\RequirePackage{fix-cm}
\documentclass[smallextended]{svjour3}       
\smartqed  
\usepackage{graphicx}
\usepackage{subcaption}
\usepackage{comment}
\usepackage{color}
\usepackage{CJKutf8}
\usepackage{lineno}
\usepackage{arydshln}
\usepackage{ulem}
\modulolinenumbers[1]
\usepackage{mathptmx}      
\usepackage[backend=biber, style=numeric, sorting=none, defernumbers=true]{biblatex}
\begin{document}

\title{
A Deep Learning Approach to Improving the Arrival-Direction Accuracy of Gamma-Ray Air-Shower Measurements with the Tibet-III Air-Shower Array
}
\titlerunning{
A Method for Improving the Arrival Direction of Gamma-Ray Showers
}        
\author{
S.~Okukawa$^{1}$,
M.~Kobayashi$^{1}$,
K.~Hara$^{2}$,
K.~Hibino$^{3}$,
Y.~Katayose$^{4}$,
K.~Kawata$^{5}$,
M.~Ohnishi$^{5}$,
T.~K.~Sako$^{6}$,
T.~Sako$^{5}$,
A.~Shiomi$^{2}$
and M.~Takita$^5$
}
\authorrunning{S. Okukawa et al.} 
\institute{M. Kobayashi \\
                \email{kobayashi-masato-fx@ynu.jp}        \\
1 Graduate School of Engineering Science, Yokohama National University, Yokohama, Kanagawa 240-8501, Japan\\
2 College of Industrial Technology, Nihon University, Narashino 275-8576, Japan\\
3 Faculty of Engineering, Kanagawa University, Yokohama, Kanagawa 221-8686, Japan\\
4 Faculty of Engineering, Yokohama National University, Yokohama, Kanagawa 240-8501, Japan\\
5 Institute for Cosmic Ray Research, University of Tokyo, Kashiwa, Chiba 277-8582, Japan\\
}
\date{Received: date / Accepted: date}
%
%
\maketitle
%
%
\begin{abstract}
The Tibet AS$\gamma$ experiment observes cosmic gamma rays from several teraelectron volts to the petaelectron volt range using 
an array of ground-based surface air-shower detectors (Tibet-III), 
which is composed of plastic scintillators. 
The arrival direction is conventionally reconstructed by fitting the shower front to detector hit-timing data, and the resulting angular resolution is approximately 0.5$^\circ$ to 0.2$^\circ$ in the energy range from 10 to 100 TeV.
In this study, to further improve the directional reconstruction accuracy, we developed a new arrival-direction reconstruction method that integrates a convolutional neural network (CNN) with the conventional method. 
Evaluations using gamma-ray events generated by Monte Carlo simulations show that the angular resolution is improved by approximately 
10\,\%--15\,\% compared with that achieved with the conventional method. In addition, 
this performance improvement shows little dependence on the zenith angle up to 40$^\circ$. 
The proposed method provides a performance gain equivalent to increasing the Tibet-III array, which consists of approximately 600 detectors, 
by roughly 150 to 200 detectors.
\end{abstract}
\section{Introduction}
\label{intro}
\quad In recent years, the air-shower observation technique using surface detector arrays has advanced significantly. 
Through observations of gamma-ray-induced air showers with this type of instrument, the Tibet AS$\gamma$ collaboration was 
the first in the world to reveal the existence of gamma rays with energies greater than 100 TeV from the Crab Nebula~\cite{Tibet_PRL2019}. This collaboration has achieved many important results, including observations of Galactic diffuse gamma rays in the sub-petaelectron volt region~\cite{Tibet_PRL2021} and the discovery of PeVatron candidates~\cite{Tibet_Nature2021}. 
The Milagro, the High Altitude Water Cherenkov(HAWC), and the Large High Altitude Air Shower Observatory(LHAASO) experiments have also reported many high-energy gamma-ray sources with energies greater than several tens of teraelectron volts~\cite{Milagro2007,Milagro2008,HAWC_science,HAWC_APJ2024,LHAASO_nature2021}.
These observations indicate that the galaxy contains numerous sources with harder spectra than previously expected and suggest that galactic cosmic-rays are accelerated beyond 1\,PeV. 
Various models have been proposed to explain charged particle acceleration up to the petaelectron volt region, but no definitive conclusion has been reached~\cite{KeFang2024,Wilhelmi2024}.
To address this issue further, observation projects using surface detector arrays in the Southern Hemisphere are underway to study sub-petaelectron volt gamma-ray sources toward the galactic center and the southern galactic plane, where many sources are concentrated. These projects include the Andes Large-area PArticle detector for Cosmic-ray physics and Astronomy (ALPACA) project~\cite{ALPAQUITA_Anzorena,ALPAQUITA_OKUKAWA,ALPAQUITA_KATO,NIMA_Ohura} and the Southern Wide-field Gamma-ray Observatory (SWGO) project~\cite{SWGO_ICRC2021}.
\par To measure air showers with high energy and angular resolution using surface detector arrays, 
it is important to measure both the lateral distribution of particle density 
and the arrival-time distribution of shower-front particles near the atmospheric depth at which the number of air-shower particles reaches its maximum.
In addition, high-statistics measurements above 10\,TeV require observations with a large effective area. 
For this reason, many experiments, including Tibet AS$\gamma$, deploy large-scale surface air-shower arrays at high-altitude sites above 4,000\,m.
\par Air-shower particles propagate at nearly the speed of light and reach the Earth’s surface while forming a conical ``shower front,'' which has a finite thickness.
Near the shower core, particles arrive in a temporally concentrated manner, producing a very sharp peak with a width of several tens of nanoseconds, whereas farther from the core the arrival times become more dispersed, with a spread of several hundred nanoseconds~\cite{Agnetta_ASTROPAR,WCD_NAKADA}. 
The shape and thickness of the time structure of the shower front are related not only to the determination of the shower-axis direction but also to the primary energy and the depth of shower maximum; they are therefore important for understanding the overall behavior of the shower~\cite{Linsley1962,TA_ICRC2019,AKENO_ICRC1985,AKENO_JPhysG1989}.
\par In surface detector array experiments, the shower axis is commonly determined using the shower front. 
In this approach, the relative times of secondary particles measured at each detector are fitted with a function that assumes a particular shower-front shape describing the spatial distribution of particle arrival times. 
The most basic function is the plane-front model, in which the shower front is approximated as a plane with a uniform inclination~\cite{AUGER_JINST,LHAASO_ChinesePhysC,Tibet2009,Tibet2010}. 
The spherical-front or curved-front model is a more realistic description, in which the shower front is treated as spherical or paraboloidal~\cite{AUGER_JINST}. 
Parametric models based on empirical formulas derived from observational data are also widely used~\cite{TA_ICRC2019,HAWC_APJ2024_2}. 
The angular resolution obtained with these methods is known to be approximately proportional to the inverse square root of the number of measured shower particles~\cite{NORIKURA1990,POIRIER1987,WCD_NAKADA}. 
Therefore, irrespective of which function is used, achieving high angular resolution requires not only precise timing measurements of secondary particles but also an increase in the number of measured particles.
To achieve improved performance, the HAWC group constructed a dense air-shower array consisting of large water-Cherenkov detectors, each approximately 4\,m high and 7.3\,m in diameter, at an altitude of 4,100\,m.
This configuration lowered the energy threshold to several hundred gigaelectron volts and achieved an angular resolution of approximately 0.1$^\circ$ above 10 TeV, defined as the angular radius containing 68\,\% of events~\cite{HAWC_APJ2024_2,HAWC1,HAWC2}.
The Astrophysical Radiation with Ground-based Observatory at Yangbajing (ARGO-YBJ) group carried out an observation experiment with a dense array of resistive plate chambers covering an area of approximately 5,800 m$^2$ 
at an altitude of 4,300\,m. 
Because of the high detector density, the energy threshold was reduced to a few hundred gigaelectron volts and the angular resolution was approximately 0.4$^\circ$ at 5 TeV, 
defined as the angular radius containing approximately 72\,\% of events~\cite{ARGO1,ARGO2,ARGO3}. 
The LHAASO group deployed the KM2A array, consisting of approximately 5,200 scintillation detectors distributed over 1.3\,km$^2$. 
Its angular resolution, defined as the angular radius containing 68\,\% of events, reaches 0.5$^\circ$--0.8$^\circ$ at 20\,TeV and 0.24$^\circ$--0.3$^\circ$ 
at 100\,TeV~\cite{LHAASO_ChinesePhysC}.
\par By contrast, the Tibet AS$\gamma$ group uses an array of plastic scintillators equipped with 5-mm-thick lead plates. 
In this experiment, approximately 600 scintillation detectors are arranged on a 7.5\,m grid at an altitude of 4,300\,m 
to continuously observe air showers induced by cosmic rays and gamma rays with energies greater than several teraelectron volts. 
The lead plate, with a thickness of one radiation length, converts secondary gamma rays in the air shower into electrons, 
thereby approximately doubling the scintillation light proportional to the energy deposited in the scintillator 
and improving the angular resolution by a factor of approximately $\sqrt{2}$. 
As a result, the angular resolution, defined as the angular radius containing 50\,\% of events, 
was improved to approximately 0.5$^\circ$ at 10\,TeV and approximately 0.2$^\circ$ at 100\,TeV~\cite{Tibet_PRL2019,Tibet2009,Tibet2010} .
\par In the present study, we developed a shower-axis reconstruction method for gamma-ray-induced air showers observed by the Tibet AS$\gamma$  experiment by combining the conventional shower-front fitting approach with a machine-learning algorithm. 
We generated simulated experimental data using Monte Carlo (MC) simulations of the air shower and the detector responses and then used this data to create 
two-dimensional images composed of detector positions, detected particle-density data, and timing data.
We fed this data into a convolutional neural network (CNN), which excels at image recognition, to train a feature extractor.
Finally, we trained an arrival-direction estimator using both the results from the shower-front fit and the features obtained from the CNN.
Section\ref{sec:experiment} describes the Tibet AS$\gamma$ experimental apparatus. 
Section\ref{sec:sim_setting} describes the gamma-ray-induced air-shower simulations and detector-response simulations. 
Section\ref{sec:air_shower_rec} describes the conventional shower reconstruction and event-selection methods. 
Section\ref{sec:combined} describes the construction of a CNN-based feature extractor that takes image file format data as input, as well as a combined-analysis method that integrates these features with the results of the shower-front fit.
Section\ref{sec:results} presents the results of applying this combined-analysis method to gamma-ray events from the Crab Nebula generated via MC simulations and evaluates its performance.
%
%
%
\section{Tibet AS$\gamma$ experiment}  
\label{sec:experiment}
\begin{figure}[htbp]
 \begin{minipage}{0.5\hsize}
 \vspace{2.5cm}
 \begin{center}
\includegraphics[width=5cm,pagebox=mediabox]{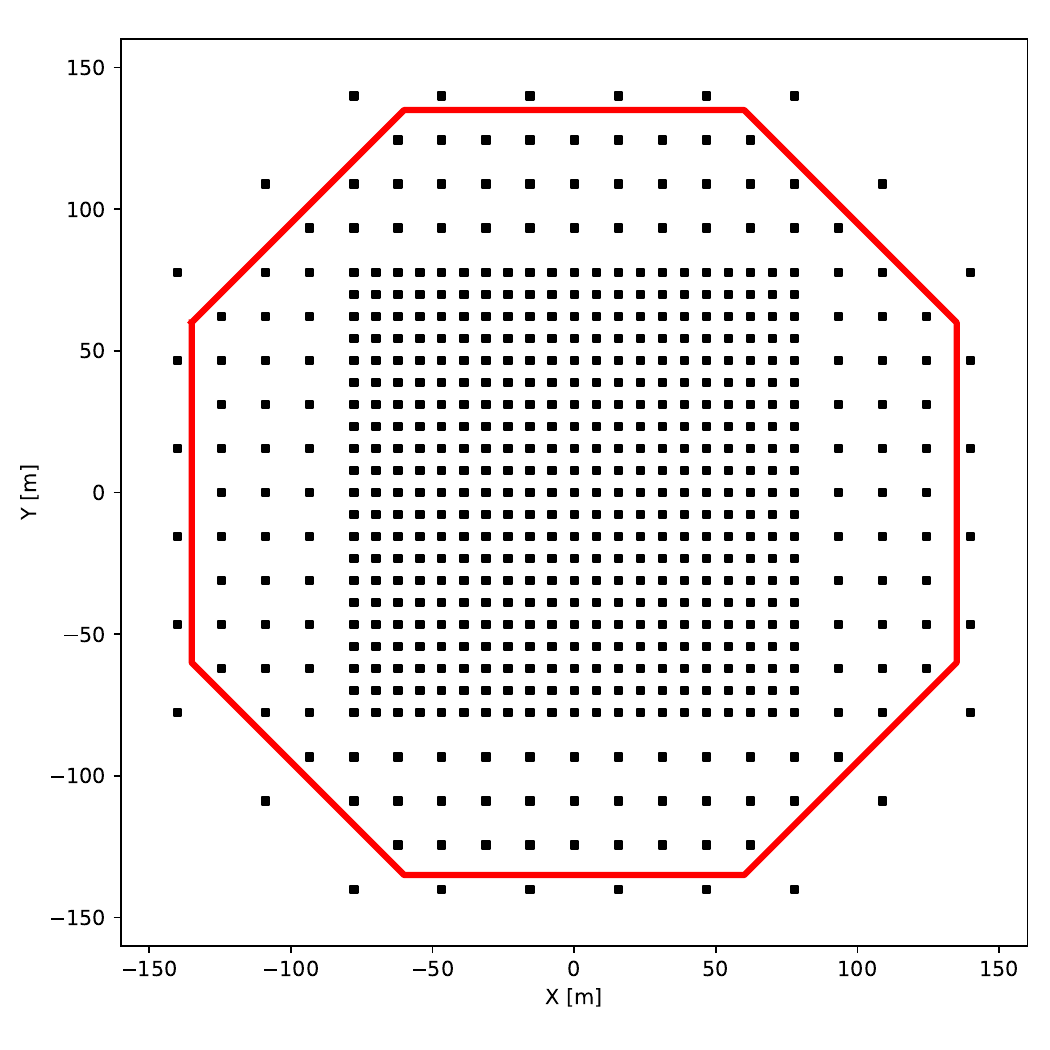}
\subcaption{\label{Fig:TibetIII}Tibet-III air-shower array.}
  \end{center}
 \end{minipage}
 \begin{minipage}{0.5\hsize}
 \vspace{2.5cm}
 \begin{center}
\includegraphics[width=7cm,pagebox=mediabox]{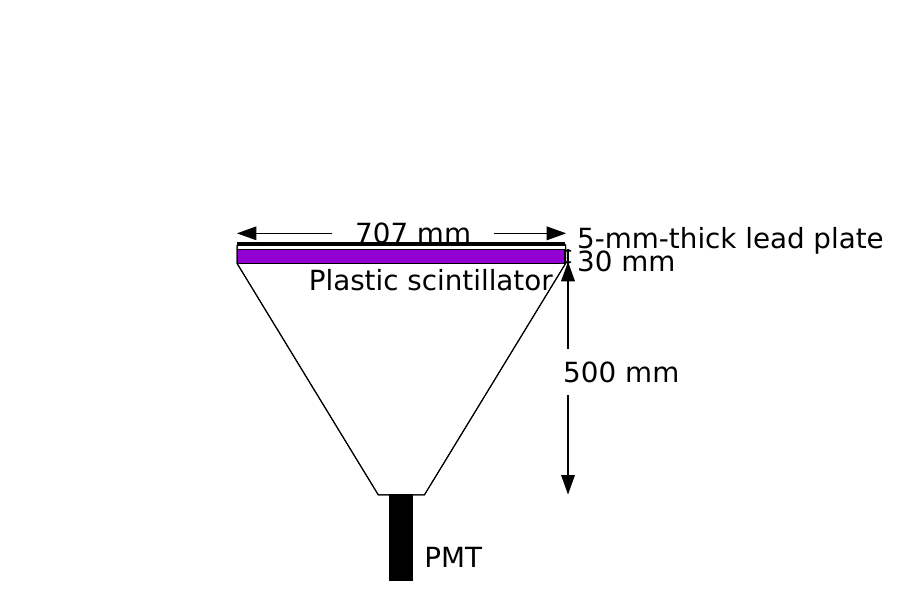}
\subcaption{\label{Fig:ASdetecoto}A plastic scintillation detector.}
  \end{center}
  \end{minipage}
\vspace{1cm}
 \caption{\label{fig:Tibet-III}
The left figure shows a top view of the Tibet-III air-shower array. The black square indicates a plastic scintillation detector with an area of 0.5 m$^2$. 
The red solid line indicates the area of the inner detector.
The figure on the right shows the structure and size of each detector.
}
\end{figure}

\quad The Tibet AS$\gamma$ experiment is located at Yangbajing on the Tibetan Plateau in China 
($90.522^{\circ}$ east, $30.102^{\circ}$ north, 4,300\,m a.s.l., 
606 g/cm$^2$ atmospheric depth)
and observes air showers induced by gamma rays and cosmic rays with energies greater than several teraelectron volts~\cite{Tibet_PRL2019,Tibet_PRL2021,Tibet_Nature2021}. 
The detector system consists of the surface air-shower array (Tibet-III) and the underground muon detector array (MD). 
Tibet-III consists of 597 scintillation detectors (Fig.~\ref{fig:Tibet-III}(a)), covering an area of approximately  65,700\,m$^2$. 
Each detector consists of a 30-mm-thick plastic scintillator topped with a 5-mm-thick lead plate and has an area of 0.5\,m$^2$ (Fig.~\ref{fig:Tibet-III}(b)). 
Each detector measures the energy loss of air-shower secondary charged particles and gamma rays converted into electrons by the lead plate; 
this quantity is converted into the particle density $\rho$ [particles m$^{-2}$]. 
The sum of particle densities, $\Sigma \rho$, is proportional to the energy of the primary particle; in this study, we used  $\Sigma \rho$ to determine the energy of primary gamma rays.
The shower arrival direction was reconstructed by fitting the shower front, derived from the hit timing at each detector, with a conical surface~\cite{Tibet_APJ2008,ALPAQUITA_KATO}. 
The energy resolution for 10, 100, and 400\,TeV gamma rays is approximately 40\,\%, 20\,\%, and 10\,\%, respectively\cite{Tibet_PRL2019,Tibet_PRL2021}, 
and the angular resolution, defined as the angular radius containing 50\,\% of events, is approximately 0.5$^\circ$ and 0.2$^\circ$ for 10 and 100\,TeV gamma rays, respectively~\cite{Tibet_PRL2019}.
\par At a depth of 2.4\,m below Tibet-III, water-Cherenkov-type muon detectors are installed. 
Each MD cell is a concrete water tank with an area of 7.35$\,\mbox{m}\times$7.35\,m and a depth 1.5\, m, 
with a downward-facing 20-inch photomultiplier tube mounted on the ceiling. 
Each cell is filled with water, and its inner walls are covered with white Tyvek sheets~\cite{TibetMD_SAKO}. 
This inner surface functions not only as a waterproof layer but also as a reflector for Cherenkov photons. 
The number of muons in an air shower measured by the MD is approximately 0--1 for 10 TeV gamma rays (compared with approximately 24 for background cosmic rays) 
and approximately 2 for 100\,TeV gamma rays (compared with approximately 180 for background cosmic rays). By applying cuts based on the number of muons measured by the MD, the cosmic-ray background can be reduced to 1.1\,\% while retaining 70\,\% of gamma-ray events above 10\,TeV, and 99.92\,\% of cosmic rays can be rejected above 100\,TeV~\cite{Tibet_PRL2019}. 
Therefore, in this study, we investigated an arrival-direction reconstruction method for pure gamma-ray-induced air showers.
\section{Air-shower simulation and detector simulation}  
\label{sec:sim_setting}
\quad Gamma-ray-induced air showers were generated using CORSIKA 7.6400~\cite{CORSIKA} in the energy range from 300 GeV to 10 PeV, with incident zenith angles smaller than 60$^\circ$.
For the combined-analysis method using the CNN described in Section~\ref{sec:combined}, 
events were generated separately for training and testing. 
Table~\ref{tab:2} shows the simulation conditions.
For the training data, showers were generated with a power-law energy spectrum of $E^{-2}$ and with isotropic arrival directions to secure sufficient statistics in the high-energy region and to avoid introducing bias into the learning of the correspondence between shower images and arrival directions. 
For the test data, gamma-ray showers from the direction of the Crab Nebula were assumed with an energy spectrum proportional to $E^{-2.6}$.
To eliminate systematic bias caused by the geometrical arrangement of the detectors, shower cores were randomly distributed 
within a radius of 300\,m from the center of the array. 
Secondary electrons and gamma rays were traced to 1\,MeV, muons to 50\,MeV, and hadrons to 1\,GeV, 
and the secondary-particle information at an altitude of 4,300\,m was recorded. 
For the Tibet-III detector configuration, a 5-mm-thick lead plate was placed above a plastic scintillator with an area of 0.5m$^2$, 
and 597 detectors were arranged at 7.5\,m intervals (Fig.~\ref{fig:Tibet-III}(a)). 
GEANT4.10.02 was used to calculate the energy loss of secondary air-shower particles in the plastic scintillator and their arrival timing. 
These quantities were converted into digital values and stored in the same format as the experimental data.
%
%
\begin{table}[h]
\caption{Simulation settings}
\label{tab:2}       
\begin{tabular}{lll}
\hline\noalign{\smallskip}
 purpose & Training & Test \\ 
 \hline
 Primaries   	                     & \multicolumn{2}{c}{Gamma-rays}  \\
 Energy regions                     & \multicolumn{2}{c}{300 GeV -- 10 PeV}\\
 Energy spectra of primaries & $\frac{dN}{dE} \propto  E^{-2}$    &  $\frac{dN}{dE} \propto  E^{-2.6}$  \\
 Arrival direction of primaries   & All directions uniform &  Crab Nebula orbit   \\
 Zenith angle                        & \multicolumn{2}{c}{$< 60$ deg.}  \\
 Observation altitude         &  \multicolumn{2}{c}{4,300\,m}  \\
 Core position of air shower  &  \multicolumn{2}{c}{ $<$ 300\,m from the center of Tibet-III}  \\
  Minimum kinetic energy &    \multicolumn{2}{c}{electrons, gamma-rays, $\pi^{0}$: 1\,MeV} \\
   of secondaries         &  \multicolumn{2}{c}{hadrons: 1\,GeV, muons : 50\,MeV} \\
 Total number of primaries    &  $9.9969 \times 10^7$\,events  &  $6.888 \times 10^8$\,events   \\
 Air-shower simulation code & \multicolumn{2}{c}{ CORSIKA7.6400} \\
 Detector simulation code & \multicolumn{2}{c}{ GEANT4.10.02} \\
 \noalign{\smallskip}\hline
\end{tabular}
\end{table}
\section{Initial analysis for air-shower reconstruction}
\label{sec:air_shower_rec}
\quad The Tibet AS$\gamma$ experiment uses traditional methods to estimate the primary energy and arrival direction using the pulse charge and hit timing measured by each scintillation detector in Tibet-III.
The pulse charge measured by each scintillation detector is converted into the particle number density $\rho$[/m$^2$]. 
The total density summed over all detectors, $\Sigma\rho$, is correlated with the energy of the primary cosmic gamma ray or hadronic cosmic ray and is therefore used for energy estimation~\cite{Tibet2009,Tibet_KURASHIGE}. 
At energies greater than approximately 100\,TeV, fitting the lateral distribution of shower particle density with the Nishimura--Kamata--Greisen (NKG) function~\cite{NKG} yields better accuracy and is therefore used in the high-energy region~\cite{Tibet_APJ2008}.
Furthermore, for primary gamma-ray showers, the quantity known as ``S50,'' which is derived from an application of the NKG-function fit, is effective~\cite{Tibet_KAWATA}. 
In this study, the energy of gamma-ray showers was determined using $\rho$ 
and the data were divided into bins with logarithmic intervals in $\Sigma\rho$, to which the new arrival-direction reconstruction method described 
in Section~\ref{sec:combined}---specifically, the combined-analysis method---was applied.
\subsection{Well reconstructed event selection}
\quad First, to use well-reconstructed shower events from the total detected data, we applied the following selection criteria.
\begin{enumerate}
    \item At least four detectors must be hit within a coincidence time window of 600\,ns.
    \item At least four detectors with $\rho \ge 0.6$ must be included among the inner detectors of the array.
\end{enumerate}
Here, the inner detectors are those located in the region enclosed by the red solid line in Fig.~\ref{fig:Tibet-III}.
\begin{enumerate}
    \setcounter{enumi}{2}
    \item At least five of the six detectors with the highest detected particle densities must be contained within the inner-detector region.
     \item The residual error from the air-shower front, defined using the cone-fit method, must be less than 1.0\,m.
    \item The zenith angle reconstructed by the cone-fit method must be less than $40^\circ$.
\end{enumerate}
The analysis described below was carried out using shower events that satisfied these selection criteria.
%
%
\subsection{Reconstruction of air-shower direction using the cone-fit method} 
 \label{subsec:reconstruction}
\begin{figure}[htbp]
\includegraphics[width=10cm,pagebox=mediabox]{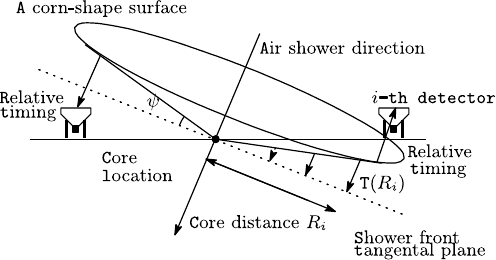}
 \caption{
Description of the cone-fit method applied to the air-shower front. The slope of the cone is defined as a function
of the distance from the obtained core location $R_i$ (m) as $T(R_i) = \frac{\tan{\psi}}{c} R_i$. }
\label{Fig:showerfit}
 \end{figure}

\quad Conventionally, shower-front fitting has been used in many experiments to determine the shower-axis direction~\cite{ALPAQUITA_KATO,NORIKURA1990,WCD_NAKADA,Tibet_APJ2008}.
The Tibet AS$\gamma$ experiment uses the ``cone-fit method'' based on the assumption of a conical shower front.
As shown in Fig.~\ref{Fig:showerfit}, if the detection time at the $i$-th detector is denoted by $t'_i$, 
then the time $t_i$ at the position where the detector is projected onto the plane perpendicular 
to the shower axis is first calculated as
\begin{equation}
t_i = t'_i - T(R_i).
\end{equation}
Here, $T(R_i)$ is the correction time that takes into account the cone slope $\psi$ and is given by
\begin{equation}
	T(R_i) = \frac{\tan{\psi}}{c} R_i. 
\end{equation}
Here, $R_i$ is the distance from the shower core to the $i$-th detector, 
where the core position is calculated as the weighted average of the detector position vectors $(x_i,y_i)$ using the particle density $\rho_i$
measured at each detector. 
The quantity $c$ is the speed of light. The true direction-cosine vector of the shower axis $(l,m,n)$ and the detector position vector $(x_i,y_i, z_i)$
satisfy
\begin{equation}
 lx_i + my_i + nz_i = -c(t_i -t_0).
\end{equation}
Therefore, the most probable direction-cosine vector can be determined by minimizing
\begin{equation}
\chi ^2 = \sum_i^{}{ w_i\left[ lx_i + my_i + nz_i + c(t_i -t_0) \right ]^2}.
\label{eq:direc_chi2}
\end{equation}
Here, $t_0$ is the time at which the shower-fitting plane passes through the origin of the detector coordinate system
and $w_i$ is the weight obtained by normalizing the number of particles detected by each detector, $w_i= \rho _i /\sum_{i}^{}{\rho _i}$.
The shower arrival direction obtained with this method was used both as the initial value for the combined method and as the reference for comparison with the combined method.
%
%
%
%
\section{Combined analysis using reconstructed air-shower parameters and CNN features from RGB image file format data }
\label{sec:combined}
\begin{figure}[htbp]
\includegraphics[width=10cm,pagebox=mediabox]{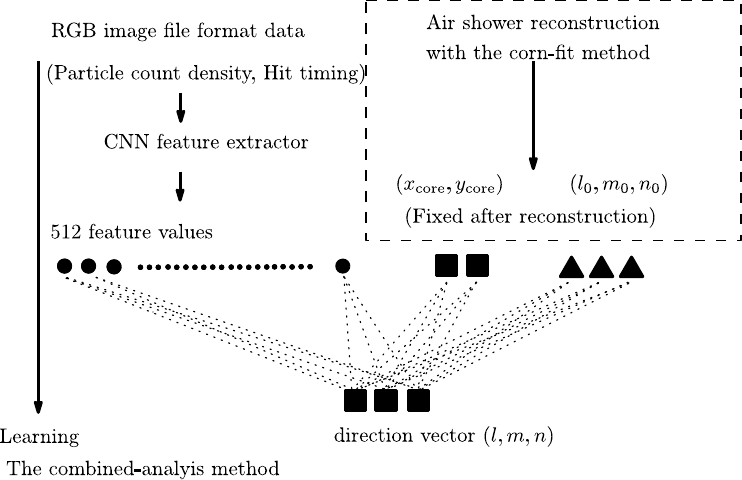}
 \caption{
The combined-analysis flow using the feature extractor of a CNN and the initial air-shower reconstruction analysis.
First, we estimated the initial results from air-shower reconstruction, consisting of the shower core $ \left ( x_{\mathrm{core}}, y_{\mathrm{core}} \right )$ and the directional vector of the shower axial $\left (l_0,m_0,m_0 \right)$.
Second, we trained a CNN feature extractor with 512 outputs and a perceptron at the same time.
In the CNN, we used RGB image file format data containing particle number density information.
The input to the final layer of the perceptron consists of the five initial results from the shower reconstruction and 512 feature values from the CNN.
}
  \label{fig:CombinedAna}
  \end{figure}

\quad In this study, to improve the angular resolution of gamma-ray-induced air showers, 
we developed a machine-learning method that uses both numerical data obtained from the conventional shower-reconstruction procedure used in the Tibet AS$\gamma$ experiment and RGB image file format data generated from shower particle densities and detector hit timings. 
The analysis consists of two main parts: first, an image-feature extraction part using a CNN, and second, a shower-parameter calculation part based on the conventional air-shower reconstruction method (Fig.~\ref{fig:CombinedAna}). 
The features extracted by the CNN were combined with the direction cosines and core position obtained from the shower-reconstruction method as initial values, and learning was performed through fully connected layers.
\subsection{Design of RGB image file format data} 
\label{subsec:image}
\begin{figure}[htbp]
 \begin{minipage}{0.5\hsize}
 \vspace{2.5cm}
 \begin{center}
\includegraphics[width=6cm,pagebox=mediabox]{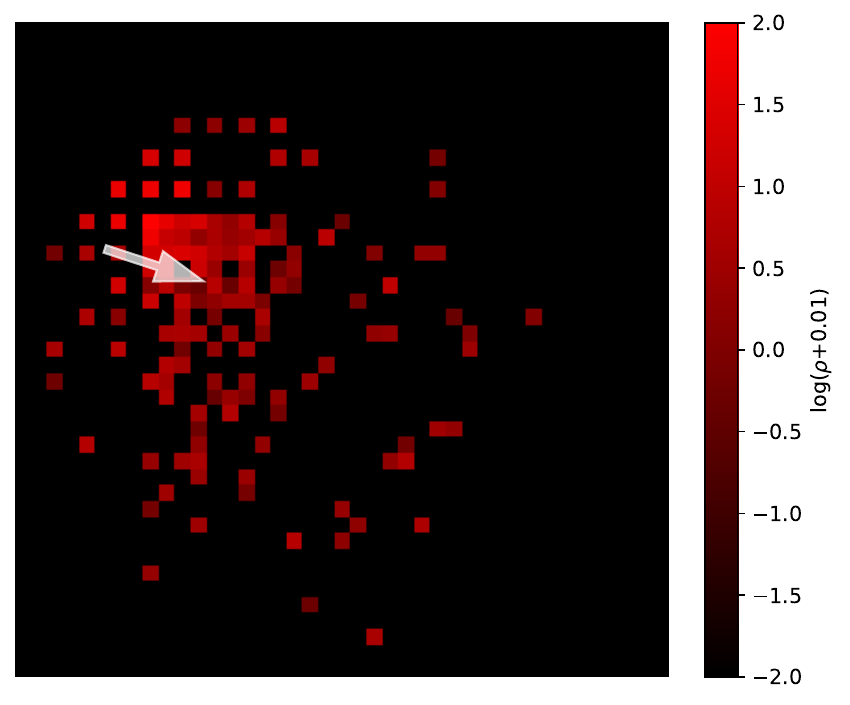}
\subcaption{\label{Fig:SampleMap_rho}Particle count density map. }
  \end{center}
 \end{minipage}
 \begin{minipage}{0.5\hsize}
 \vspace{2.5cm}
 \begin{center}
\includegraphics[width=6cm,pagebox=mediabox]{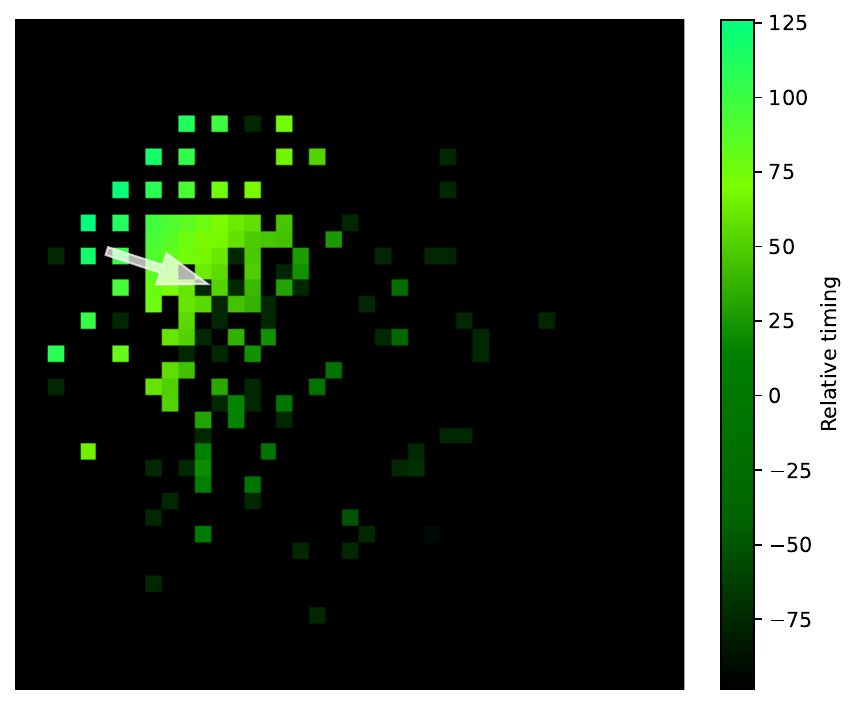}
\subcaption{\label{Fig:SampleMap_hittiming}Hit timing map. }
  \end{center}
  \end{minipage}
\vspace{1cm}
 \caption{\label{Fig:SampleMap}
 Example of particle number density map and a hit timing map.
 RGB image file format data contains information on particle number density and hit timing.
These figures display results for the primary gamma-ray energy of 98.6\,TeV.
The square pixel on the left shows the number density, $\log{\left (\rho + 0.01 \right )}$, of a detector at that location, and the square pixel on the right shows the relative value of detector hit timing.
}
 \end{figure}

\quad This analysis used a CNN that accepts RGB color image input. 
First, we converted the particle-density data and hit-timing data obtained from 597 detectors 
arranged on a 7.5\,m grid into RGB image file format data. 
Specifically, we divided a region of 307.5\,m $\times $ 307.5\,m 
into a 41 $\times 41$ grid and stored the detected particle number density $\rho$ 
and detector hit time $t_i'$ in each cell. 
Cells without detectors were assigned values of zero. 
By storing particle density and time in two of the three RGB channels and setting the remaining channel to 0, we created a $41\times41\times3$ data array.
Fig.~\ref{Fig:SampleMap} shows an example of the re-visualization of this array data for a 98.6\,TeV gamma-ray shower.
Fig.~\ref{Fig:SampleMap_rho} shows a particle number density map, with the horizontal and vertical axes corresponding to array coordinates. 
The color of each cell represents the logarithm of the particle number density, $\log{(\rho + 0.01)}$, 
and the white cross indicates the reconstructed shower core position. 
The arrow indicates the direction of the shower-axis obtained from the initial analysis described in Section~\ref{sec:air_shower_rec}.
In this example, a high-density region is formed near the shower core, and the density decreases toward the periphery.
The color distribution in the timing map is generally consistent with the shower-axis direction estimated in the initial analysis.
\subsection{Combined-analysis method }
\label{subsec:CombAna}
%
\par
\quad We used EfficientNetB1~\cite{Tan2019EfficientNetRM} as the CNN-based feature extractor. 
EfficientNetB1 consists of seven blocks followed by Conv2D, BatchNormalization, and Activation layers, and its standard input size is 
$240 \times 240 \times 3$ (RGB). 
Because the data size is reduced with each processing step in each analysis block,
we enlarged the original RGB image file format data of size $41 \times 41 \times 3$ to $246 \times 246 \times 3$ 
to avoid excessive degradation of information due to over-reduction. 
During this enlargement, each original value was simply replicated into the corresponding expanded elements. 
The output of EfficientNetB1 is converted into a 1,280-dimensional feature vector by a Global Average Pooling layer and then further compressed to 512 dimensions, which is the default for EfficientNetB1, via a fully connected layer.
Dropout with a rate of 0.4 was subsequently applied for regularization.
\par This 512-dimensional vector was concatenated with the core position $(x_{\mathrm{core}}, y_{\mathrm{core}})$ obtained 
from the shower reconstruction and the arrival-direction cosines $(l_0, m_0, n_0)$ determined by the cone-fit method described in Subsection~\ref{subsec:reconstruction} (Fig.~\ref{fig:CombinedAna}). 
These five values were fixed during training. 
The initial weights for the 512 features and $(x_{\mathrm{core}}, y_{\mathrm{core}})$ were 
set according to a truncated normal distribution with mean 0 and standard deviation $\sigma$ randomly chosen in the range $10^{-4} \leq \sigma \leq 10^{-3}$. By contrast, the weights connecting $l_0$ to $l$, $m_0$ to $m$, and $n_0$ to $n$ were initialized to 1. 
In this way, a final 517-dimensional input vector (= 512 + 5) was constructed, regularized with dropout, and fed into a fully connected layer that outputs the arrival-direction vector $(l, m, n)$.
Note that we trained the CNN extractor and the fully connected layer simultaneously.
\begin{table}[h]
  \caption{Outline of the architecture of feature extractor of a CNN.}
  \label{tb:architecture}
  \centering
  \begin{tabular}{| l l | c |}
    \hline
     \textbf{Procedure}      &  \begin{tabular}{c}\end{tabular}  & \textbf{Output data size}   \\
    \hline \hline
    Inputs data              & size: 246$\times$246$\times$3  & --  \\
    \hline
     EfficientNetB1          & default setting & 8$\times$8$\times$1280 \\
     \hline
     Global average pooling  & --  & 1280 \\
     \hline
     Fully-connected layer   & --  & 512 \\
     \hline
    Dropout                  & --  & 512 \\
     \hline
       \end{tabular}
\end{table}
For both training and testing, we used air-shower events that satisfy the conditions in Section~\ref{sec:air_shower_rec}
and divided the events into bins of $\Sigma \rho$ for training and testing. 
The total numbers of events were 83,785 and 97,032, respectively. 
In addition, 10\,\% of the training data were used as validation data and the validation loss was calculated at each epoch. 
During training, a mini-batch size of eight was used, and the mean squared error ($MSE$) was adopted as the loss function, with backpropagation used as the learning algorithm.
\begin{equation}
MSE  = \frac{1}{n} \sum_{i=0}^{n-1} (y_{i} - y_{i,\mathrm{true}})^2
\label{eq:MSE}
\end{equation}
Here, $n$ is the mini-batch size and $y_{i,\mathrm{true}}$ and $y_i$ represent the true arrival direction and the predicted output 
for the $i$-th sample in the mini-batch, respectively.
\section{Results}  
\label{sec:results}
\quad We applied the analysis criteria in Section~\ref{sec:air_shower_rec} to gamma-ray-induced shower events generated by MC simulation 
and selected events where $\Sigma\rho$ fell within the range $100 \le \Sigma\rho < 1000$.
For these events, we compared the directional resolution obtained using the conventional air-shower reconstruction method based on cone fitting (cone-fit method) with that obtained using a new analysis method (combined-analysis method) that 
combines this method with the machine learning technique described in Section~\ref{sec:combined}.
\subsection{Energy dependence of angle resolution}
\begin{figure}[htbp]
 \begin{minipage}{0.5\hsize}
 \vspace{0.5cm}
 \begin{center}
\includegraphics[width=5cm,pagebox=mediabox]{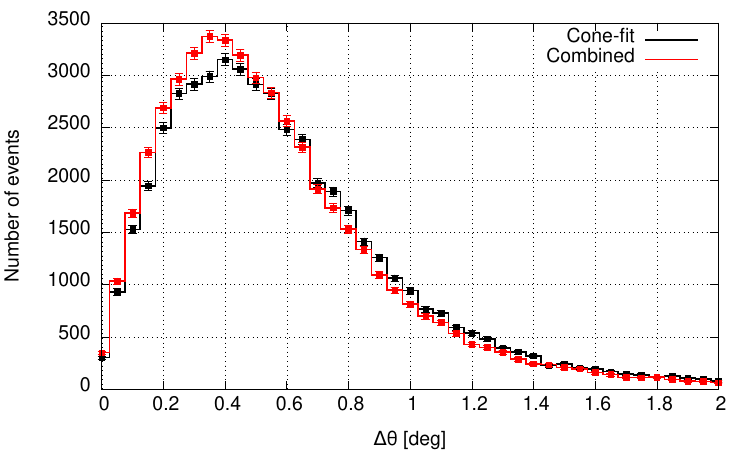}
\subcaption{$100 \le \sum \rho < 178 (E_{\gamma}=11.5\,\mathrm{TeV})$ }\label{Fig:AngDist_11.5TeV_MDcut}
  \end{center}
 \end{minipage}
 \begin{minipage}{0.5\hsize}
 \vspace{0.5cm}
 \begin{center}
\includegraphics[width=5cm,pagebox=mediabox]{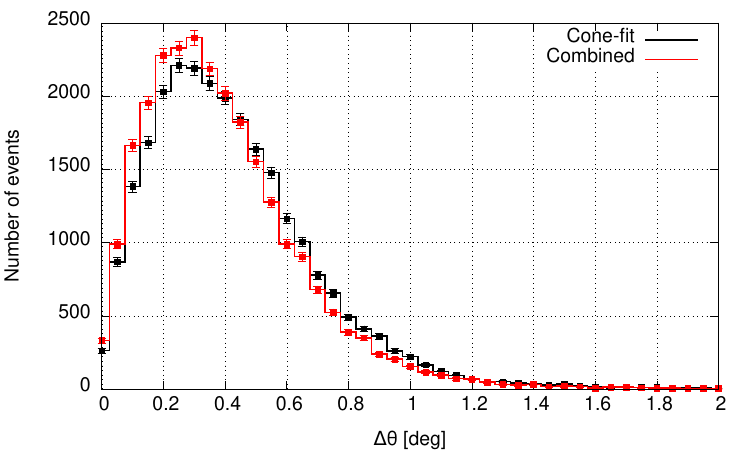}
\subcaption{$178 \le \sum \rho < 316 (E_{\gamma}=19.4\,\mathrm{TeV}) $}\label{Fig:AngDist_19.4TeV_MDcut}
  \end{center}
  \end{minipage}
 \begin{minipage}{0.5\hsize}
 \vspace{0.5cm}
 \begin{center}
\includegraphics[width=5cm,pagebox=mediabox]{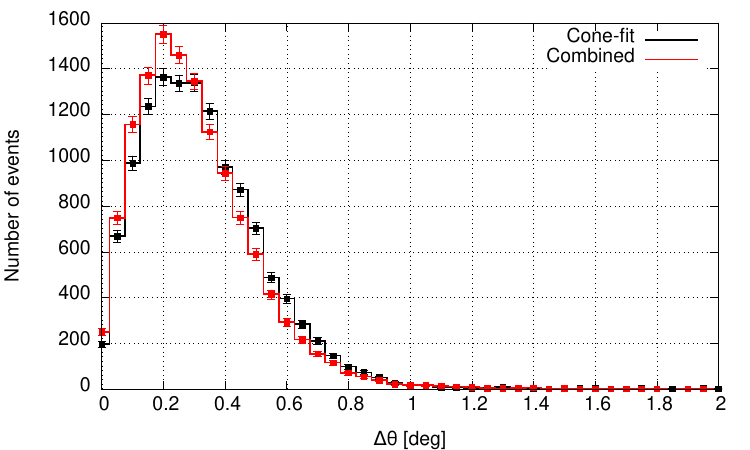}
\subcaption{$316 \le \sum \rho < 562 (E_{\gamma}=32.9\,\mathrm{TeV}) $}\label{Fig:AngDist_32.9TeV_MDcut}
  \end{center}
 \end{minipage}
 \begin{minipage}{0.5\hsize}
 \vspace{0.5cm}
 \begin{center}
\includegraphics[width=5cm,pagebox=mediabox]{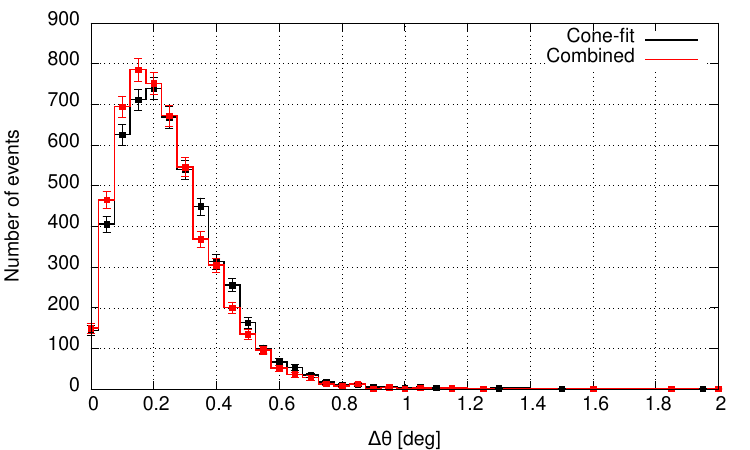}
\subcaption{$562 \le \sum \rho < 1000 (E_{\gamma}=57\,\mathrm{TeV})$}\label{Fig:AngDist_57TeV_MDcut}
  \end{center}
 \end{minipage}
  \begin{minipage}{0.5\hsize}
 \vspace{0.5cm}
 \begin{center}
\includegraphics[width=5cm,pagebox=mediabox]{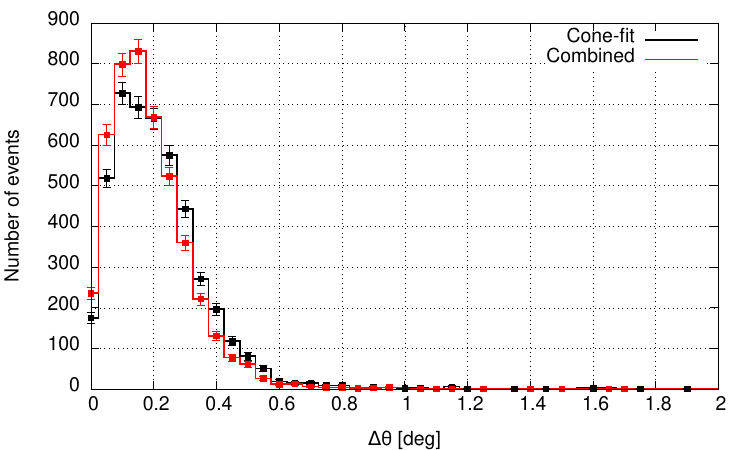}
\subcaption{$1000 \le \sum \rho < 1778 (E_{\gamma}=98.6\,\mathrm{TeV})$}\label{Fig:AngDist_98.6TeV_MDcut}
  \end{center}
 \end{minipage}
 \begin{minipage}{0.5\hsize}
 \vspace{0.5cm}
 \begin{center}
\includegraphics[width=5cm,pagebox=mediabox]{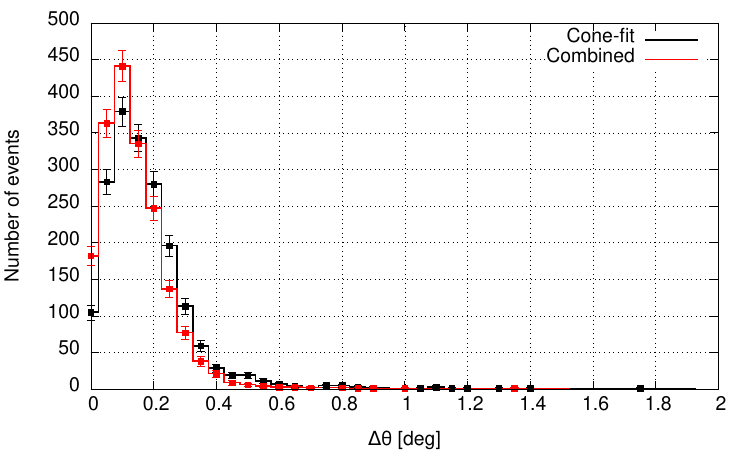}
\subcaption{$1778 \le \sum \rho < 3162 (E_{\gamma}=187\,\mathrm{TeV})$}\label{Fig:AngDist_187TeV_MDcut}
  \end{center}
 \end{minipage}
  \begin{minipage}{0.5\hsize}
 \vspace{0.5cm}
 \begin{center}
\includegraphics[width=5cm,pagebox=mediabox]{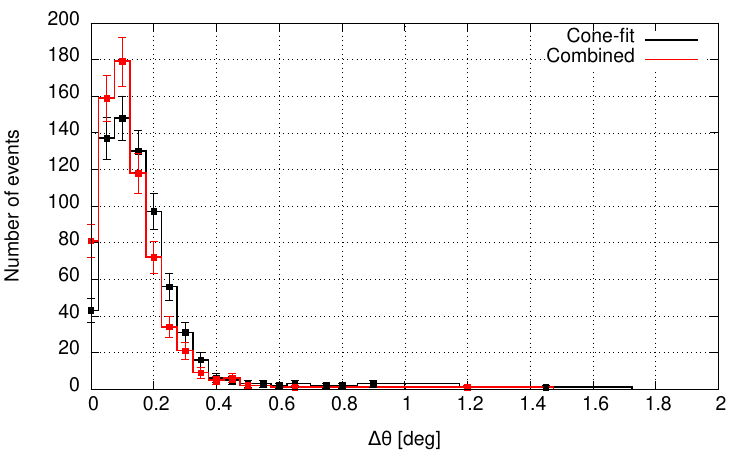}
\subcaption{$3162 \le \sum \rho < 5623 (E_{\gamma}=371\,\mathrm{TeV})$}\label{Fig:AngDist_371TeV_MDcut}
  \end{center}
 \end{minipage}
 \begin{minipage}{0.5\hsize}
 \vspace{0.5cm}
 \begin{center}
\includegraphics[width=5cm,pagebox=mediabox]{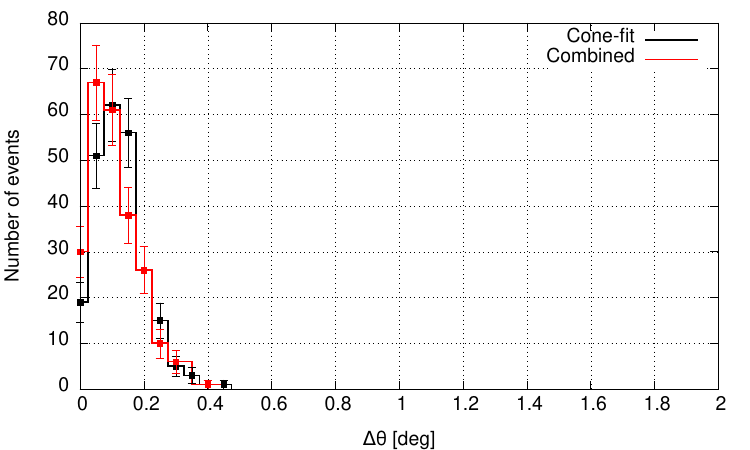}
\subcaption{$5623 \le \sum \rho < 10000 (E_{\gamma}=738\,\mathrm{TeV})$}\label{Fig:AngDist_738TeV_MDcut}
  \end{center}
 \end{minipage}
\vspace{1cm}
 \caption{\label{Fig:angres}
  The difference in the angular distance ($\Delta \theta$) between the true arrival direction and the direction calculated by the analysis.
 Figure (a)--(h) shows the result of each $\sum \rho$ bin from 100 to 10,000.
 The black histograms show the results obtained from the initial shower reconstruction, and the red histograms show the combined-analysis results.
}
 \end{figure}

\begin{figure}[htbp]
 \begin{minipage}{0.5\hsize}
 \vspace{0.5cm}
 \begin{center}
\includegraphics[width=5cm,pagebox=mediabox]{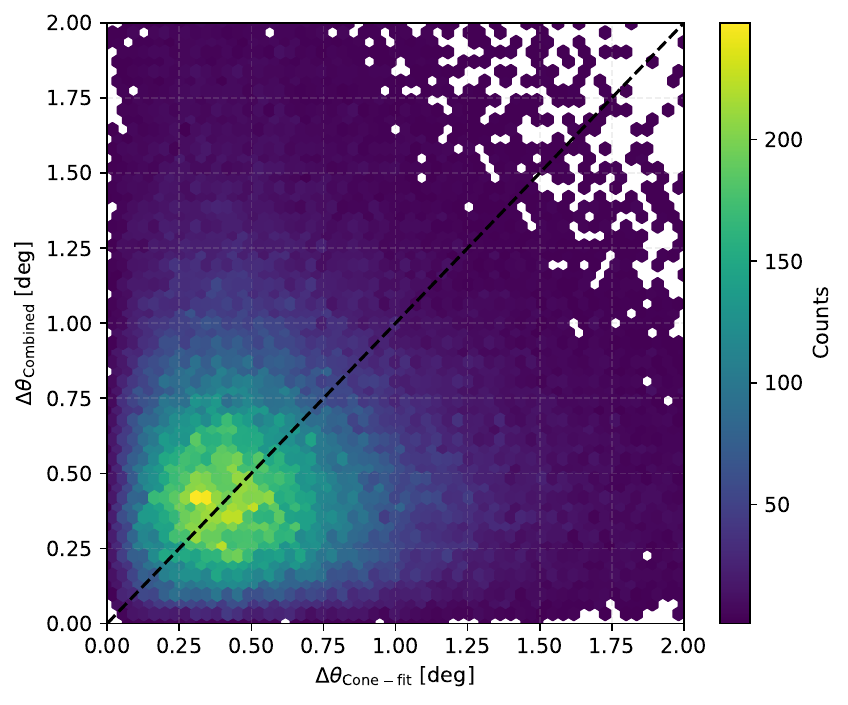}
\subcaption{$100 \le \sum \rho < 178 (E_{\gamma}=11.5\,\mathrm{TeV})$}
  \end{center}
 \end{minipage}
  \begin{minipage}{0.5\hsize}
 \vspace{0.5cm}
 \begin{center}
\includegraphics[width=5cm,pagebox=mediabox]{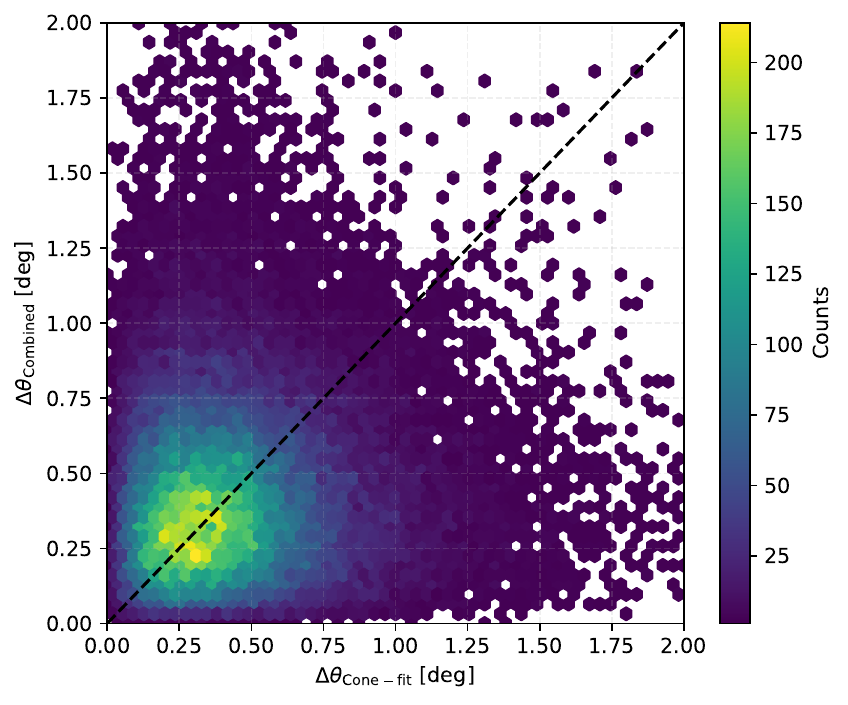}
\subcaption{$178 \le \sum \rho < 316 (E_{\gamma}=19.4\,\mathrm{TeV})$}
  \end{center}
 \end{minipage}
 \begin{minipage}{0.5\hsize}
 \vspace{0.5cm}
 \begin{center}
\includegraphics[width=5cm,pagebox=mediabox]{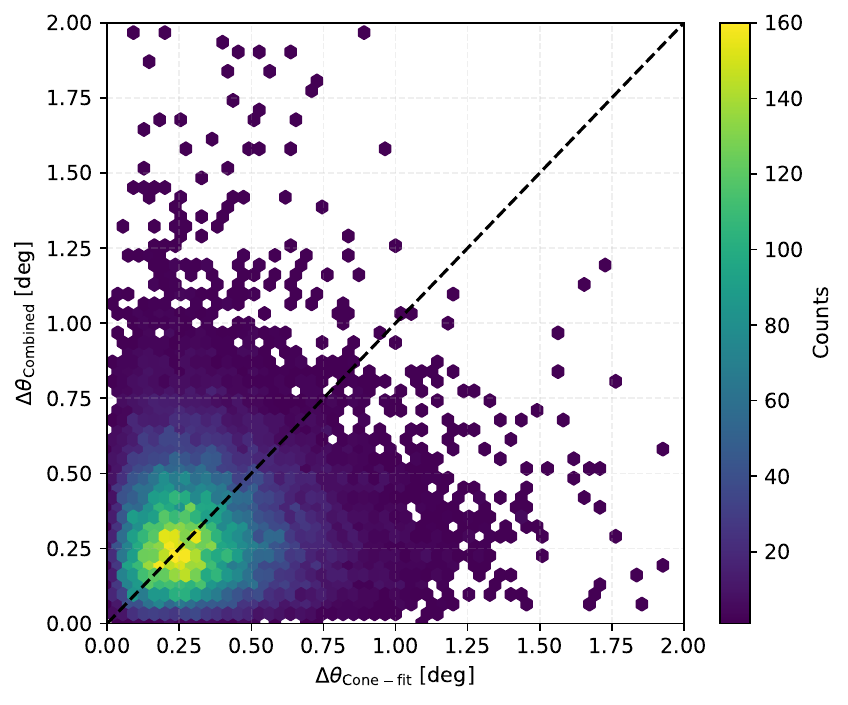}
\subcaption{$316 \le \sum \rho < 562 (E_{\gamma}=32.9\,\mathrm{TeV})$}
  \end{center}
 \end{minipage}
  \begin{minipage}{0.5\hsize}
 \vspace{0.5cm}
 \begin{center}
\includegraphics[width=5cm,pagebox=mediabox]{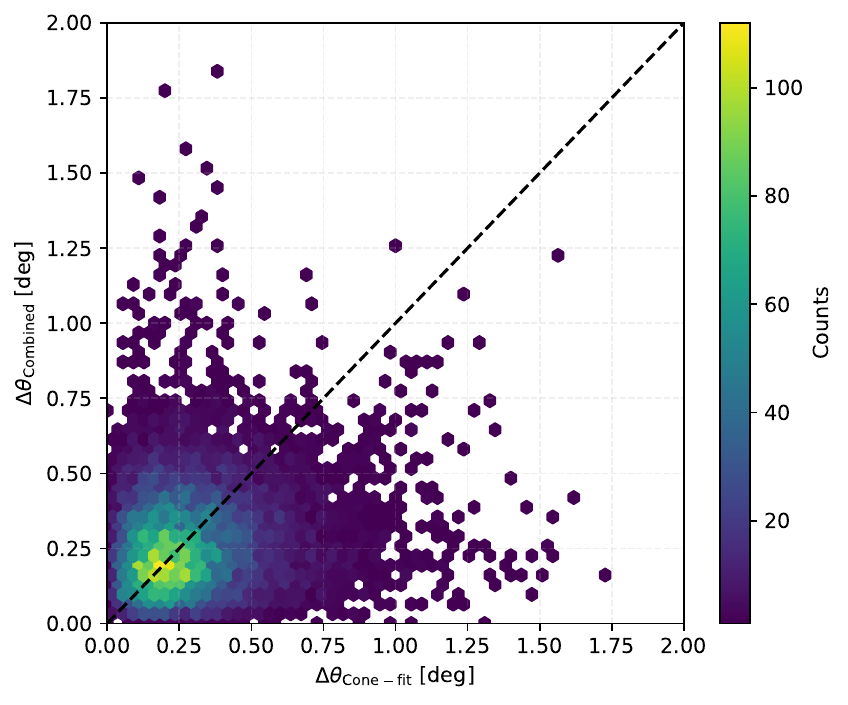}
\subcaption{$562 \le \sum \rho < 1000 (E_{\gamma}=57\,\mathrm{TeV})$}
  \end{center}
 \end{minipage}
  \begin{minipage}{0.5\hsize}
 \vspace{0.5cm}
 \begin{center}
\includegraphics[width=5cm,pagebox=mediabox]{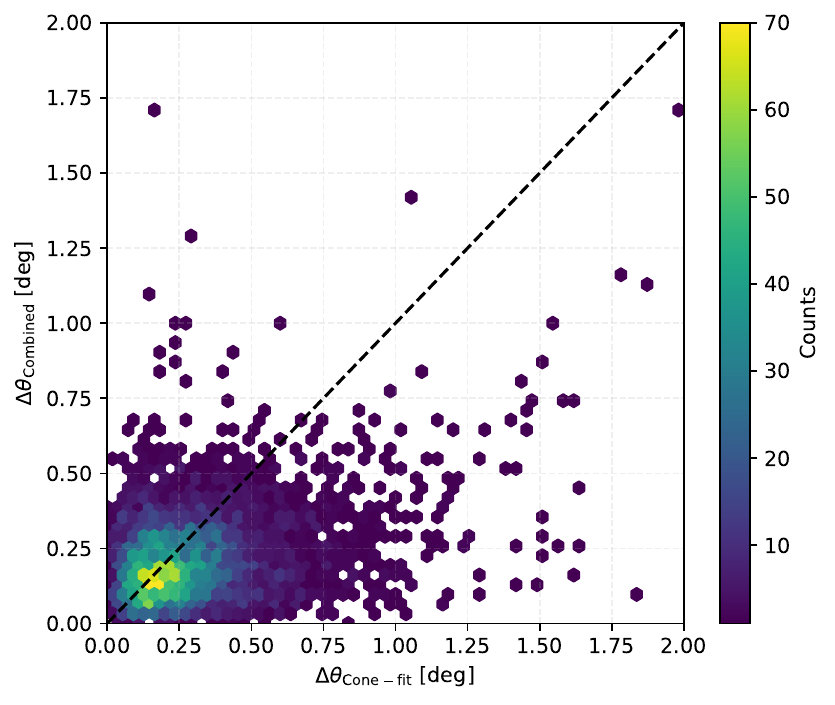}
\subcaption{$1000 \le \sum \rho < 1788 (E_{\gamma}=98.6\,\mathrm{TeV})$}
  \end{center}
 \end{minipage}
  \begin{minipage}{0.5\hsize}
 \vspace{0.5cm}
 \begin{center}
\includegraphics[width=5cm,pagebox=mediabox]{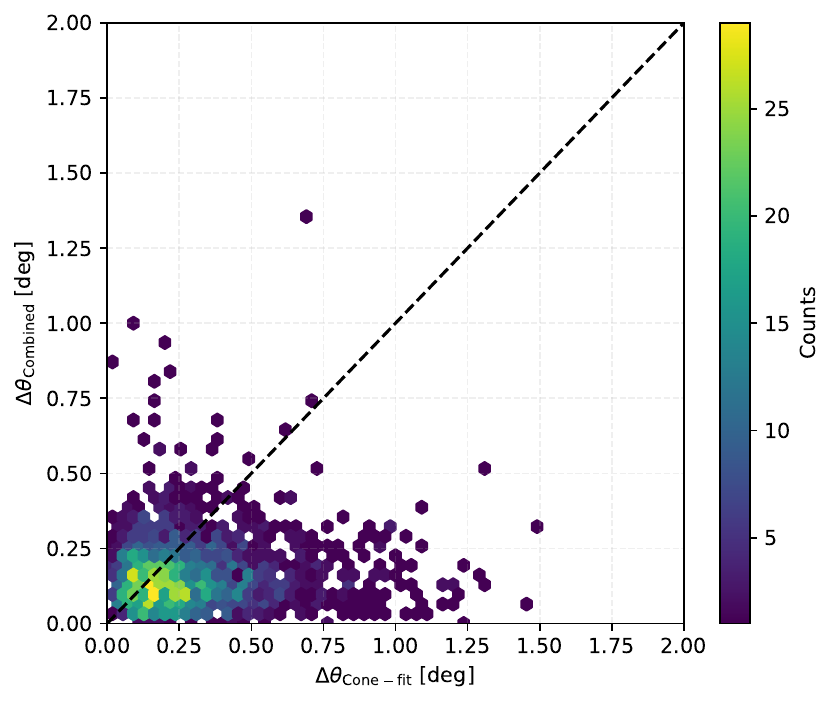}
\subcaption{$1778 \le \sum \rho < 3162 (E_{\gamma}=187\,\mathrm{TeV})$}
  \end{center}
 \end{minipage}
  \begin{minipage}{0.5\hsize}
 \vspace{0.5cm}
 \begin{center}
\includegraphics[width=5cm,pagebox=mediabox]{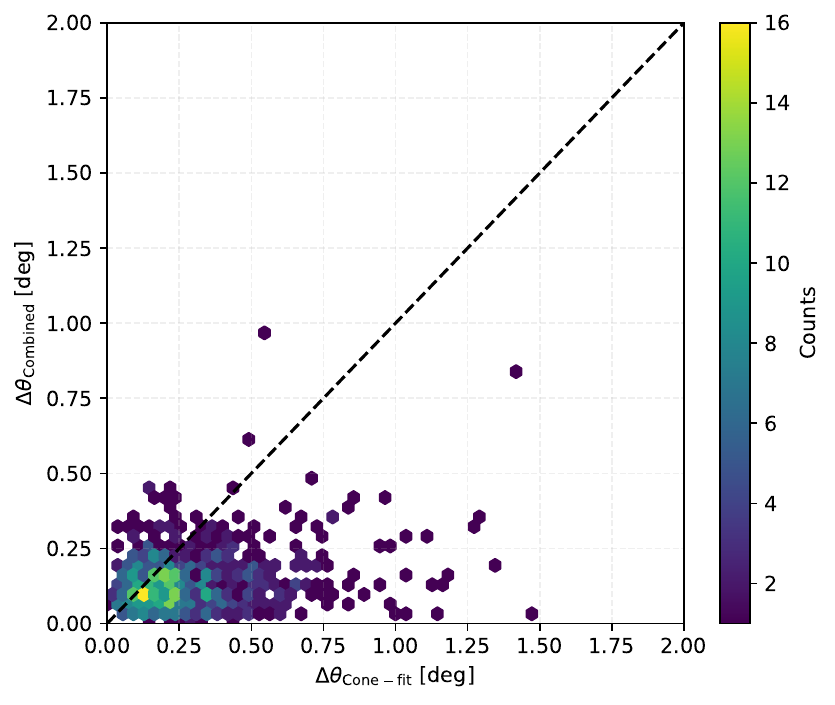}
\subcaption{$3162 \le \sum \rho < 5623 (E_{\gamma}=371\,\mathrm{TeV})$}
  \end{center}
 \end{minipage}
  \begin{minipage}{0.5\hsize}
 \vspace{0.5cm}
 \begin{center}
\includegraphics[width=5cm,pagebox=mediabox]{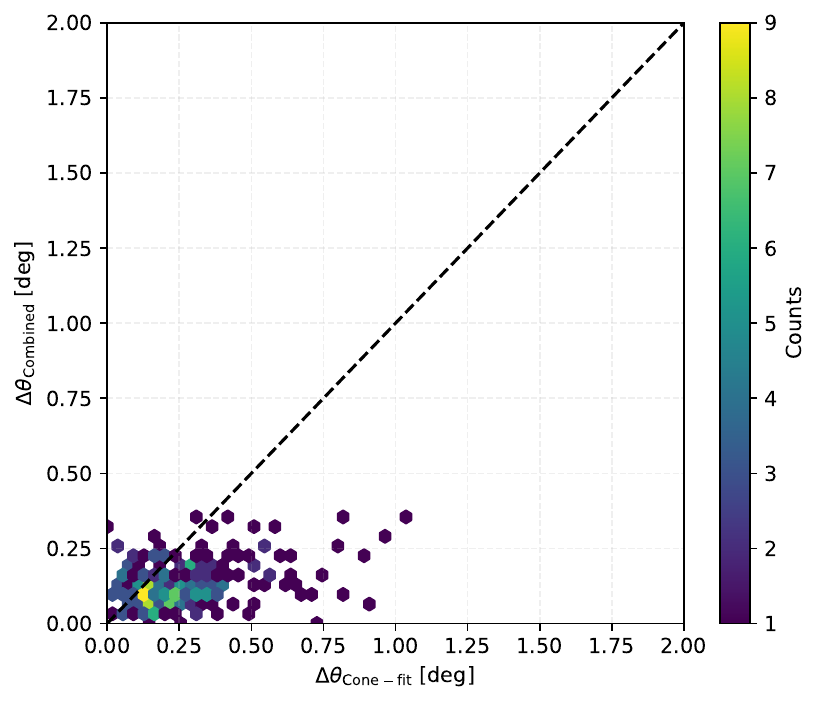}
\subcaption{$5623 \le \sum \rho < 10000 (E_{\gamma}=738\,\mathrm{TeV})$}
  \end{center}
 \end{minipage}
 \caption{
 \label{Fig:Heatmap_2res}
 Heatmaps of the angular distances calculated using the cone-fit method ($\theta_\mathrm{res\_cone\_fit} $) and the combined-analysis method ($\theta_\mathrm{res\_combined} $).
 The figures show results for eight $\sum \rho$ bins from 100 to 10,000.
}
 \end{figure}

\begin{figure}[htbp]
 \begin{minipage}{0.5\hsize}
 \vspace{0.5cm}
 \begin{center}
\includegraphics[width=5cm,pagebox=mediabox]{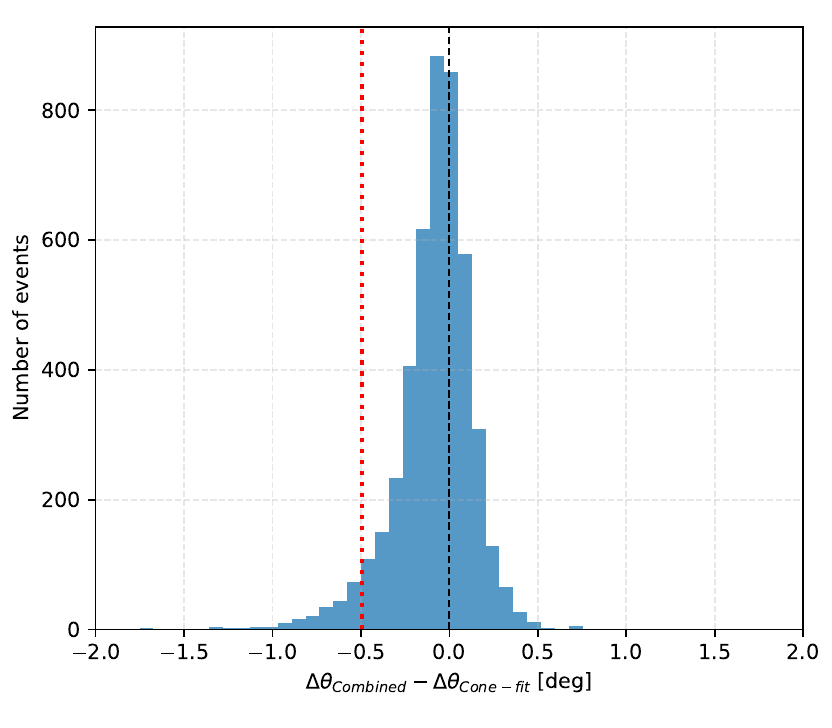}
\subcaption{Difference of two angular distances.}
  \end{center}
 \end{minipage}
  \begin{minipage}{0.5\hsize}
 \vspace{0.5cm}
 \begin{center}
\includegraphics[width=5cm,pagebox=mediabox]{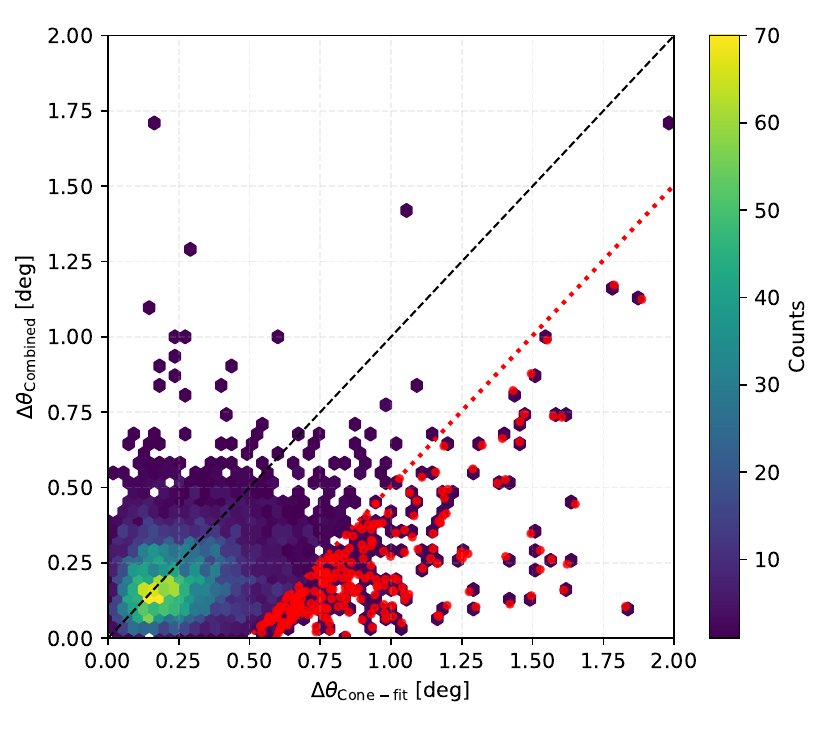}
\subcaption{Heatmaps of the angular distances calculated using two methods.}
  \end{center}
 \end{minipage}
 \begin{minipage}{0.5\hsize}
 \vspace{0.5cm}
 \begin{center}
\includegraphics[width=5cm,pagebox=mediabox]{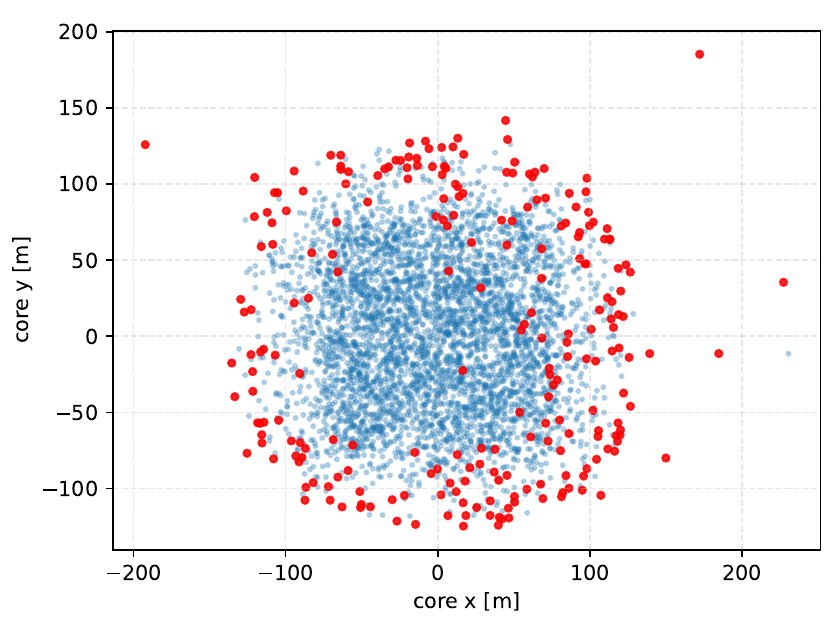}
\subcaption{Core position of air showers.}
  \end{center}
 \end{minipage}
  \begin{minipage}{0.5\hsize}
 \vspace{0.5cm}
 \begin{center}
\includegraphics[width=5cm,pagebox=mediabox]{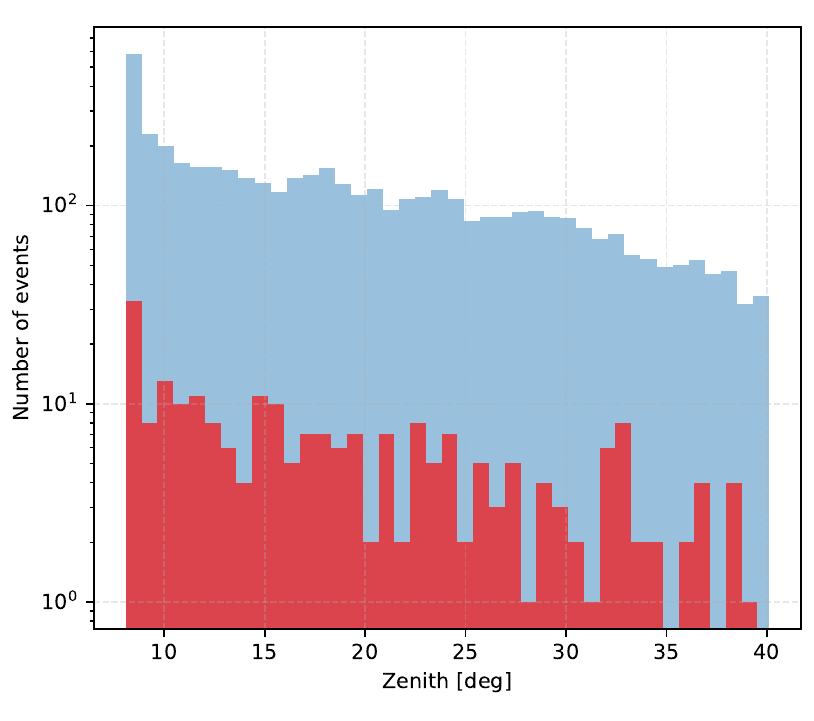}
\subcaption{Zenith angle distribution of the air shower. Core position of air showers.}
  \end{center}
 \end{minipage}
\vspace{1cm}
 \caption{\label{Fig:TwoMethod98_6TeV}
 Features of the significantly improved events with the combined-analysis method.
 Figure (a)--(d) shows the result of $\sum \rho$ bin from 1000 to 1,778.
Figure (a) shows the difference in angular distance calculated by two methods.
The red dotted line indicates the region that encompasses the bottom 5\,\% of events when the distribution is integrated from the left.
Figure (b) is the same as the lower-left figure in Figure~\ref{Fig:Heatmap_2res}, and the red dots indicate 5\,\% events.
Red points in Figure (c) indicate the core position of the 5\,\% events. 
Figure (d) shows the zenith angle distribution. 
The blue histogram is the result for all events, and the red histogram is the result for the 5\,\% event.
}
 \end{figure}

\quad Fig.~\ref{Fig:angres} shows the angular resolution for showers with true zenith angles between 0$^{\circ}$ and 40$^{\circ}$. 
The horizontal axis shows the angular distance $\Delta \theta$ between the true shower direction and the direction obtained by the analysis. 
The events are divided into eight $\Sigma\rho$ intervals; the representative energy corresponding to each interval is shown in the caption. 
The black squares represent results obtained using the cone-fit method, whereas the red circles represent results obtained using the combined-analysis method.
As the energy increases, the peak position shifts to the left and the width becomes narrower, 
indicating that the angular resolution improves with increasing energy. 
In addition, the distribution obtained with the combined-analysis method is narrower and has more events on the left-hand side than that obtained with the cone-fit method, indicating that the combined-analysis method yields better resolution.
We defined the angular resolution $\theta_\mathrm{res}$ as the angle within which 50\,\% of the total events 
are contained when integrating the event counts from angular distance 0. 
In the lowest-$\Sigma\rho$ bin, the angular resolution is approximately 0.55$^{\circ}$ for the cone-fit method and approximately 0.51$^{\circ}$ for the combined-analysis method.
In the highest-$\Sigma\rho$ bin, the corresponding values are approximately 0.14$^{\circ}$ and approximately 0.12$^{\circ}$, respectively.\\
To compare the angular distances calculated using the two methods (i.e., the cone-fit method and the combined-analysis method) for each event,
we created a heatmap for each event by plotting the angular distances ($\Delta \theta$) determined by each method on the $ x$- and $ y$-axes, respectively.
Fig. ~\ref{Fig:Heatmap_2res} shows the results for four regions where $\Sigma\rho$ ranges from 100 to 10,000.
The red line in the figure is a straight line with a slope of 1, and the color indicates the frequency of events.
Events are concentrated along the straight line, and no unexpected local distortions or biases are observed; however,
even in the low $ \Sigma \rho $ region, events from the cone-fit method appear to be more widely distributed.
As energy increases, the results from the combined-analysis method clearly show that events with large angular distances become fewer and are concentrated in a relatively small region.
To investigate which event arrival directions were improved by the combined-analysis method, we calculated 
$\Delta \theta_\mathrm{res\_combined} - \Delta\theta_\mathrm{res\_cone\_fit}$ for the results where $\Sigma\rho$ was between 1000 and 1778 in Fig.~\ref{Fig:Heatmap_2res} 
and extracted the events for which this value differed substantially.
Fig.~\ref{Fig:TwoMethod98_6TeV}(a) shows the distribution of this difference; 
we selected events by integrating from the left and including only those within the top 5\,\% of the distribution.
Fig.~\ref{Fig:TwoMethod98_6TeV}(b) shows the selected events plotted again in red on the heatmap.
The black line is a straight line with a slope of 1, and the red line indicates the 5\,\% threshold.
We found that the events shown in red correspond to air-shower cores that fell on the outer edge of the array (Fig.~\ref{Fig:TwoMethod98_6TeV}(c)).
In the cone-fit method, the shower plane is fitted and the direction is determined using the air-shower core positions obtained from the reconstruction analysis in Section~\ref{subsec:reconstruction}.
By contrast, in the combined-analysis method, image file format data, such as secondary particle density and time information, are also believed to be used for correction.
Regarding the zenith angle, when we compared the results of all events with those of the events shown in red (Fig.~\ref{Fig:TwoMethod98_6TeV}(d)), we found that no specific events had been improved.
\par
\begin{figure}[htbp]
\includegraphics[width=12cm,pagebox=mediabox]{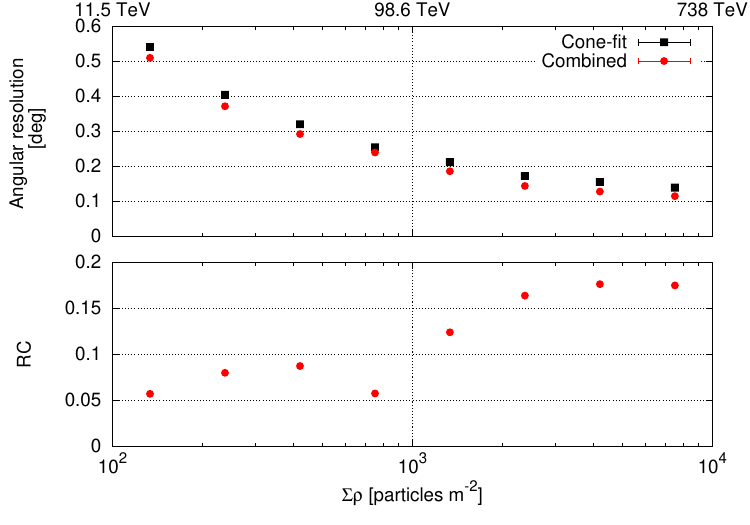}
 \caption{\label{Fig:RC_AngRes}
 Energy dependence of the angle resolution improvement.
The vertical axis of the upper figure is the angular resolution, and the horizontal axis is $\Sigma \rho$.
The black squares show values obtained from the initial shower reconstruction ($\theta_\mathrm{res\_cone-fit} $), and the red circles show the results of the combined-analysis ($\theta_\mathrm{res\_combined} $).
The lower figure shows the relative change($RC$) defined by Eq.~\ref{Eq:RC}
}
 \end{figure}

To quantify the improvement in angular resolution achieved with the combined-analysis method, we defined the relative change ($RC$) as
\begin{equation}
RC  =\frac{  \theta_\mathrm{res\_cone\_fit} - \theta_\mathrm{res\_combined} }{  \theta_\mathrm{res\_cone\_fit}} .
\label{Eq:RC}
\end{equation}
Here, $\theta_\mathrm{res\_cone\_fit}$ and $\theta_\mathrm{res\_combined}$ are the angular resolutions obtained by the cone-fit method and the combined-analysis method, respectively.
The energy dependence of the relative change is shown in the lower subfigure of Fig.~\ref{Fig:RC_AngRes};
the improvement is approximately 5.9\,\% for $100 \le \Sigma\rho < 178$(11.5\,TeV) and approximately 17\,\% for $5623  \le \Sigma\rho <  10000$(738\,TeV).
\subsection{Zenith-angle dependence of angle resolution}
\begin{figure}[htbp]
  \begin{minipage}{0.5\hsize}
 \vspace{0.5cm}
 \begin{center}
\includegraphics[width=5cm,pagebox=mediabox]{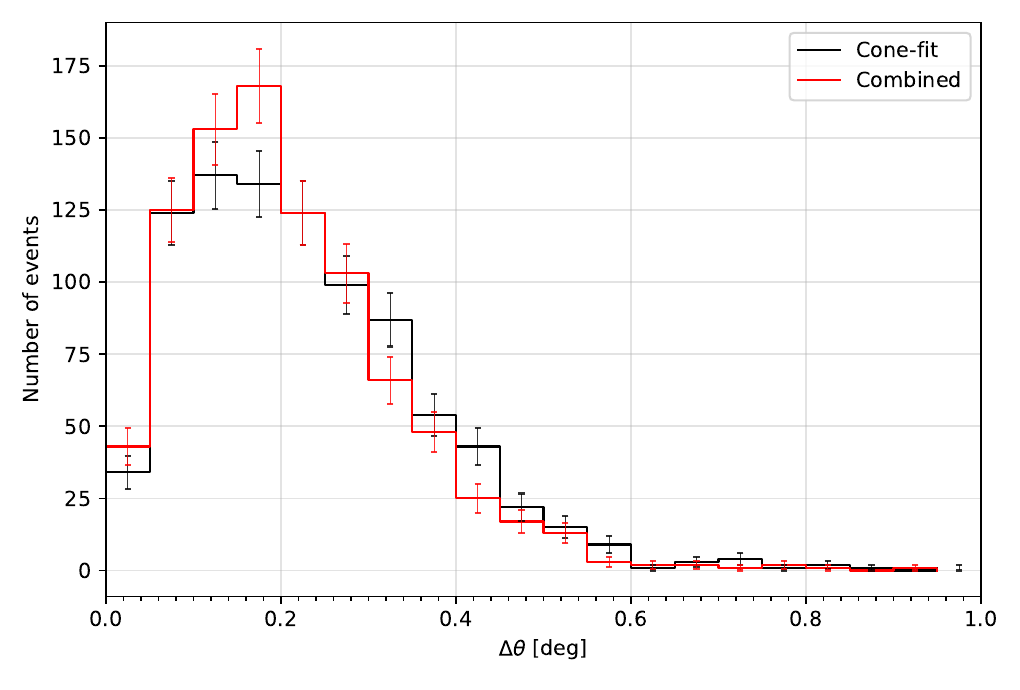}
\subcaption{$0^{\circ} \le \theta_{\mathrm{ture}} < 10^{\circ} $}\label{Fig:AngDist_98.6TeV_MDcut}
  \end{center}
 \end{minipage}
 \begin{minipage}{0.5\hsize}
 \vspace{0.5cm}
 \begin{center}
\includegraphics[width=5cm,pagebox=mediabox]{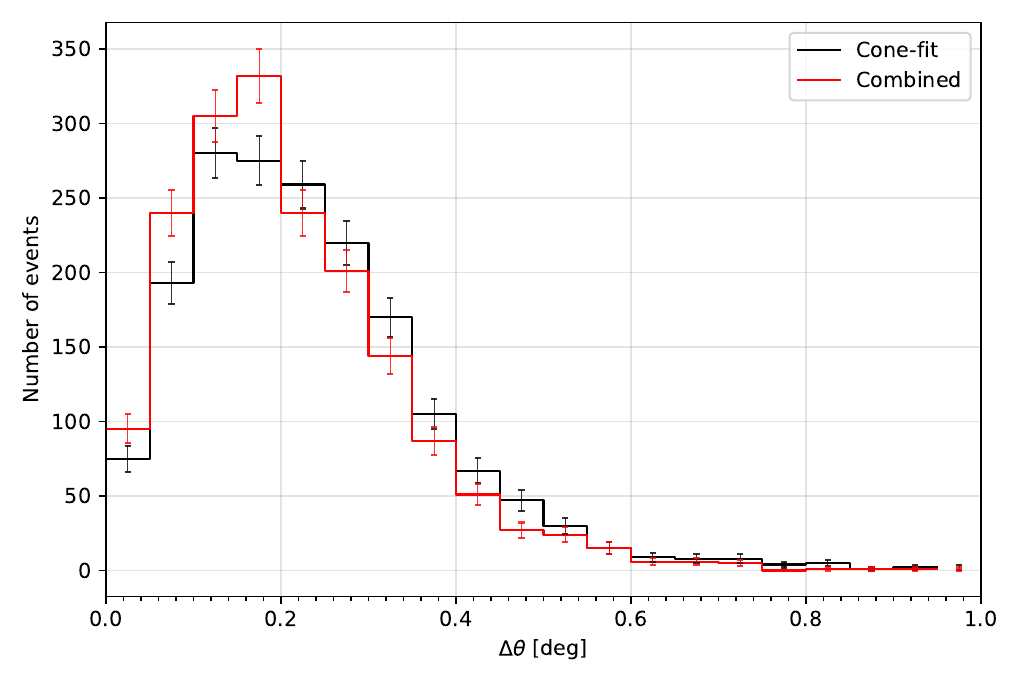}
\subcaption{$10^{\circ} \le \theta_{\mathrm{ture}} < 20^{\circ} $}\label{Fig:AngDist_187TeV_MDcut}
  \end{center}
 \end{minipage}
  \begin{minipage}{0.5\hsize}
 \vspace{0.5cm}
 \begin{center}
\includegraphics[width=5cm,pagebox=mediabox]{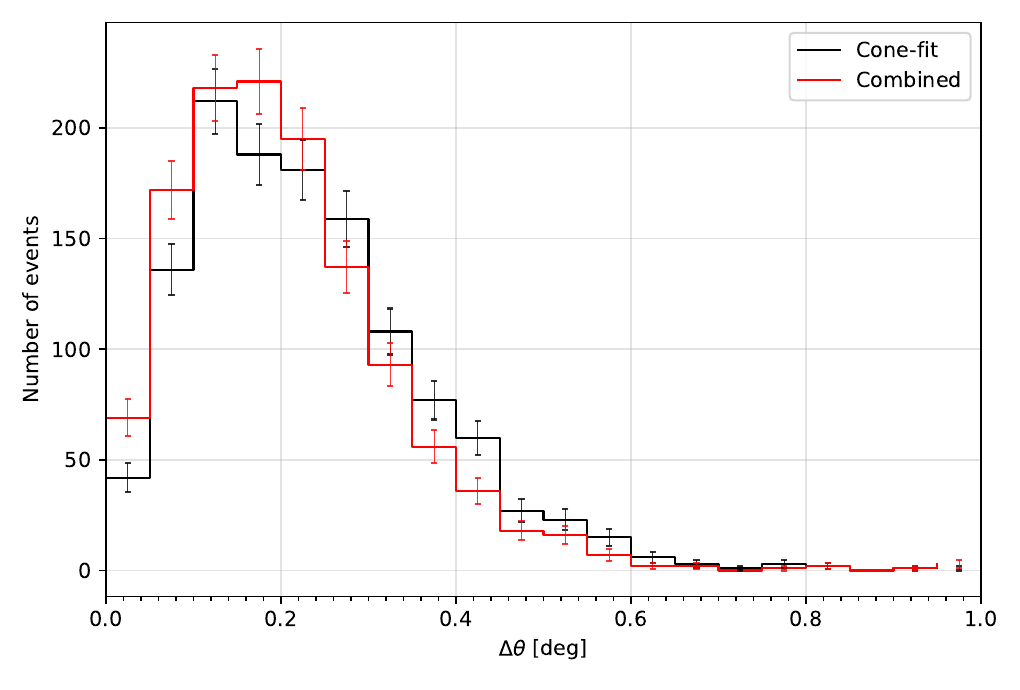}
\subcaption{$20^{\circ} \le \theta_{\mathrm{ture}} < 30^{\circ} $}\label{Fig:AngDist_371TeV_MDcut}
  \end{center}
 \end{minipage}
 \begin{minipage}{0.5\hsize}
 \vspace{0.5cm}
 \begin{center}
\includegraphics[width=5cm,pagebox=mediabox]{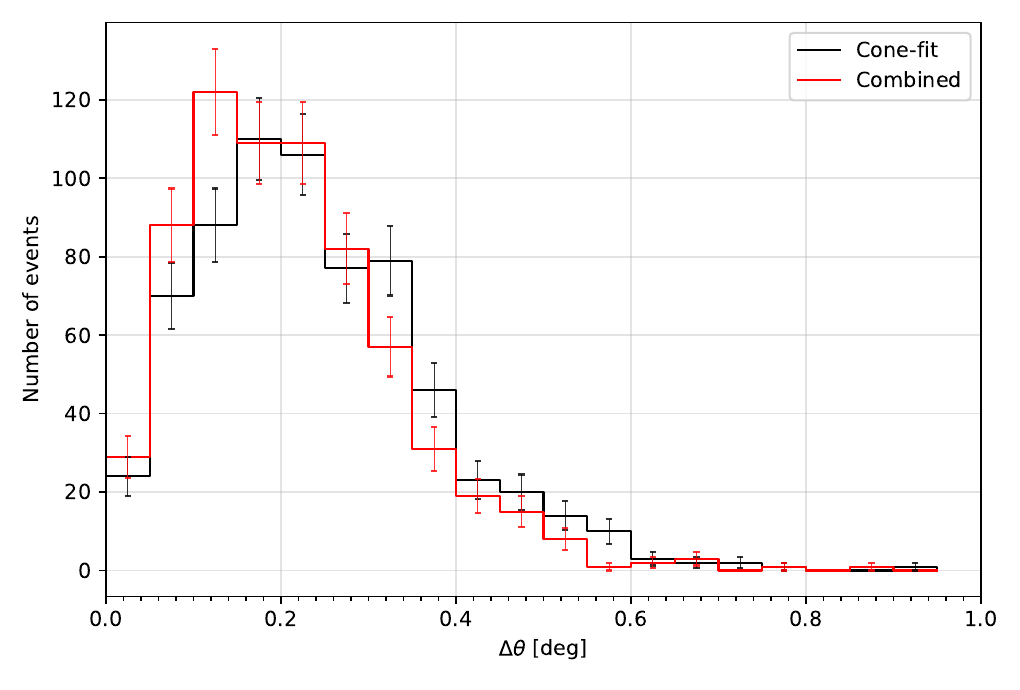}
\subcaption{$30^{\circ} \le \theta_{\mathrm{ture}} < 40^{\circ} $\label{Fig:AngDist_738TeV_MDcut} }
  \end{center}
 \end{minipage}
\vspace{1cm}
 \caption{\label{Fig:Znt_AngRes}
The angular distance between the true arrival direction ($\theta_\mathrm{ture}$) and the direction calculated by the analysis.
 Figure (a)--(d) shows the result of each actual zenith angle bin from 10$^{\circ}$ to 40$^{\circ}$.
 These figures show the results for $1000 \le \Sigma \rho < 1778$.
 The black histograms show the results obtained from the initial shower reconstruction, and the red histograms show the combined-analysis results.
}
 \end{figure}

\begin{figure}[htbp]
\includegraphics[width=12cm,pagebox=mediabox]{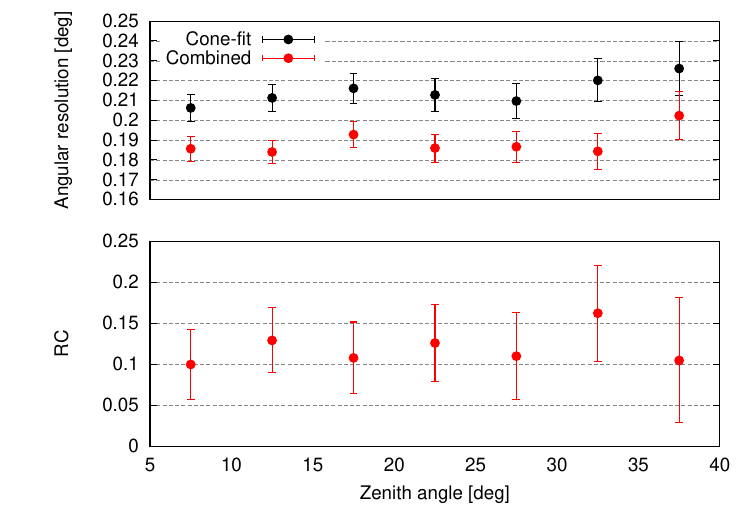}
 \caption{
 \label{Fig:RC_Znt_AngRes}
 Zenith angle dependence of angle resolution.
The vertical axis of the upper figure is angular resolution, and the horizontal axis is the true zenith angle. 
The black squares show values obtained from the cone-fit method ($\theta_\mathrm{res\_cone\_fit} $), and the red circles show the results of the combined-analysis ($\theta_\mathrm{res\_combined} $).
The lower figure shows the relative change ($RC$) defined by an equation of the same type as Equation~\ref{Eq:RC}.
}
 \end{figure}

\quad We investigated the zenith-angle dependence of the effect of the combined-analysis method. 
We examined events with true zenith angles ($\theta_\mathrm{true}$) ranging from 0$^{\circ}$ to 40$^{\circ}$ in 10$^{\circ}$ intervals; 
the angular resolution in each region is shown in Fig.~\ref{Fig:Znt_AngRes}.
This figure shows the results of events where $1000 \le \Sigma\rho < 1778$; for every zenith-angle interval, 
the distribution obtained with the combined-analysis method has slightly more events on the smaller-angle side 
and a narrower width than the distribution obtained with the cone-fit method. 
Furthermore, by dividing the zenith angle into 5$^{\circ}$ intervals, 
we investigated the dependence of the angular resolution on the zenith angle; 
the results are shown in the top subfigure in Fig.~\ref{Fig:RC_Znt_AngRes}.
The angular resolution for events with zenith angles from 0$^{\circ}$ to 10$^{\circ}$ was 0.21$^{\circ}$ 
for the cone-fit method and approximately 0.18$^{\circ}$ for the combined-analysis method. 
For events with zenith angles from 30$^{\circ}$ to 40$^{\circ}$,
the corresponding values were approximately 0.22$^{\circ}$ and 0.19$^{\circ}$, respectively. 
We calculated the relative change in angular resolution using equation ~\ref{Eq:RC} 
and plotted its dependence on the zenith angle in the bottom subfigure in Fig.~\ref{Fig:RC_Znt_AngRes}.
The improvement is approximately 10\,\%--15\,\% and is essentially independent of the zenith angle.
\subsection{Importance of features contributing to improving angular resolution}
\label{subsec:importance}
\begin{figure}[htbp]
 \begin{minipage}{1.0\hsize}
 \vspace{0.5cm}
 \begin{center}
\includegraphics[width=12cm,pagebox=mediabox]{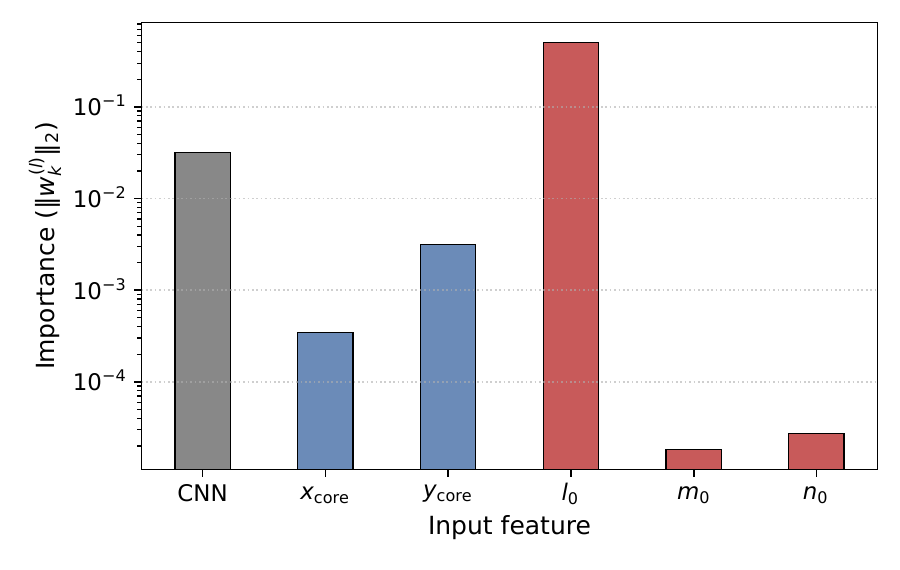}
\subcaption{Importance of six feature values. 
The ``CNN'' label indicates the summation value from the CNN's pre-training.}
  \end{center}
 \end{minipage}
 \begin{minipage}{1.0\hsize}
 \vspace{0.5cm}
 \begin{center}
\includegraphics[width=12cm,pagebox=mediabox]{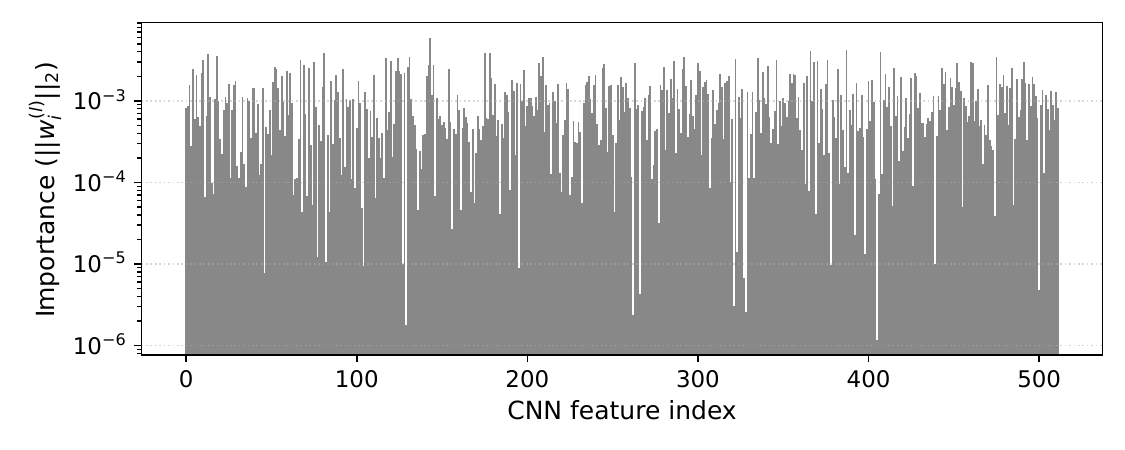}
\subcaption{Importance of 512 CNN feature values.}
  \end{center}
 \end{minipage}
\caption{
 \label{Fig:Importance}
Importance of features used in the combined-analysis method.
This shows the result for the $x$ component $(l)$ of the direction vector.
The vertical axis of the figure shows the $L2$ norm in Subsection~\ref{subsec:importance}. 
In figure (a), the vertical axis value for the ``CNN'' label in the upper
figure corresponds to the sum of the 512 features obtained via pre-training with the
CNN. Figure (b) shows the importance of each of the 512 CNN features.
}
\end{figure}

\begin{figure}[htbp]
 \begin{minipage}{0.5\hsize}
 \vspace{2.5cm}
 \begin{center}
\includegraphics[width=6cm,pagebox=mediabox]{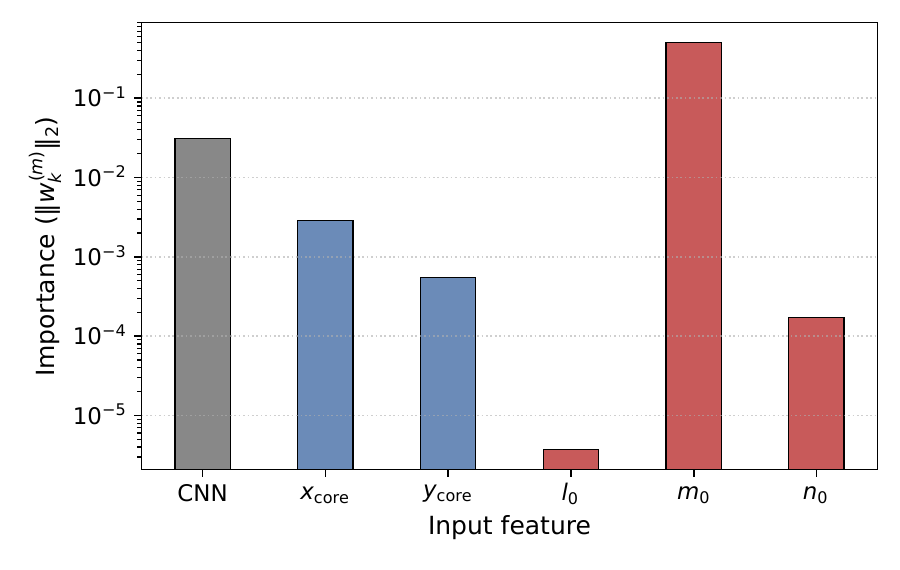}
\subcaption{Importance of features for the direction vector $m$.}
  \end{center}
 \end{minipage}
 \begin{minipage}{0.5\hsize}
 \vspace{2.5cm}
 \begin{center}
\includegraphics[width=6cm,pagebox=mediabox]{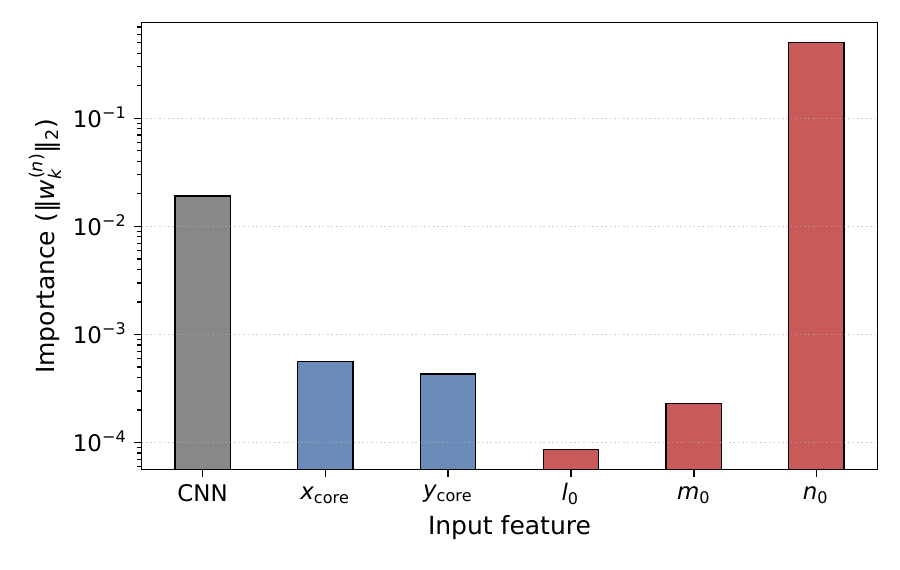}
\subcaption{Importance of features for the direction vector $n$.}
  \end{center}
  \end{minipage}
\vspace{1cm}
 \caption{\label{Fig:Importance_mn}
 Importance of features used in the combined-analysis method.
Figures (a) and (b) show the results for the $y$ component ($m$) of the direction vector and the $z$ component ($n$), respectively.
}
 \end{figure}

\quad In the combined-analysis method, to investigate the extent to which each input feature contributes to the output value, we used the following linear equation~\ref{eq:importance}, which determines the output value of the final layer.
\begin{equation}
y_j=\sum_{i=1}^{517}{w_i^{(j)}} \times x_i+b_j
\label{eq:importance}
\end{equation}
Here, $y_j$ is the final output value. The index $j$ takes values 1--3, corresponding to the direction cosines $(l,m,n)$. 
Parameter $x_i$ is an input feature, where $i$ takes values from 1 to 517, 
and consists of the five components obtained from the shower reconstruction analysis (i.e., the core position $(X_{\mathrm{core}}, y_{\mathrm{core}})$ and the direction cosines $(l_0, m_0, n_0)$), along with the features obtained from the CNN.
The 512 CNN features were summed and treated as a single feature in the evaluation of the feature importance. Therefore, the importance was evaluated for six features in total. The coefficient $w_i^{(j)}$ is the weight for output $j$, and $b_j$ is the offset. 
The importance of the $k$-th feature was defined as the $L2$ norm($  ||w_k^{(j)}||_2$) of the corresponding weight vector. 
Fig.~\ref{Fig:Importance} shows the importance of each feature used to calculate the $x$-component $(l)$ of the direction vector determined by the combined-analysis method.
The vertical axis represents the $L2$ norm.
Fig.~\ref{Fig:Importance}(a) shows the results for the six parameters,
where the contribution of the $x$-component ($l_0$) of the direction vector obtained by the cone-fit method was the largest, with $||w_k^{(l)}||_2=5.0\times10^{-1}$.
The contributions of the $y$ and $z$ components of the direction vector were small compared with that of the $x$ component, amounting to approximately $5 \times 10^{-5}$.
The second-largest contribution was from the CNN parameters, with $||w_k^{(l)}||_2 = 3.2 \times 10^{-2}$.
Furthermore, because the detector configuration is symmetric, the contribution 
from the core position $(x_{\mathrm{core}}, y_{\mathrm{core}})$ was expected to be of the same order of magnitude; however, $y_{\mathrm{core}}$ is approximately 10 times larger. We speculatively attribute this greater-than-expected contribution to fluctuations in the initial random numbers used in the analysis 
and expect that changing those numbers would alter this relationship.
The importance of each of the 512 CNN layers is shown in Fig.~\ref{Fig:Importance}(b), and all were of similar importance.
Fig.~\ref{Fig:Importance_mn} shows the results of our similar analysis of the $y$ and $z$ components of the direction cosines.
In each case, the $y$ component ($m_0$) and the $z$ component ($n_0$) contribute the most, whereas the contribution from the features extracted by the CNN is the second largest but is less than one-tenth that of $m_0$ (or $n_0$).
These trends are similar to those observed for the $x$ component.
\section{Summary} 
\label{sec:summary}
\quad To improve the angular resolution of the Tibet-III air-shower array for gamma-ray-induced air showers in the several-tens-of-teraelectron-volts region, 
we studied a new arrival-direction reconstruction method. 
The new method combines the traditional shower reconstruction technique (cone-fit method), 
which uses shower surface fits, with CNN analysis of shower image data.
In the CNN analysis, we generated RGB image file format data consisting of particle density for each detector, hit time, and detector position and extracted 512 features.
Using the cone-fit method, we determined the arrival direction vector and the shower core position.
The arrival direction was then re-estimated using a neural network that takes 
as input 517 features formed by combining the CNN features with these reconstruction parameters.
Evaluations using gamma-ray events generated by MC simulations showed 
that the angular resolution was improved by approximately 10\,\%--15\,\% relative to that achieved with the cone-fit method. 
In addition, a similar level of improvement was obtained up to zenith angles of 40 $^{\circ}$. 
The angular resolution $\theta_{\mathrm{res}} $ obtained with the cone-fit method is known to approximately scale as 
$
 \theta_{\mathrm{res}} \propto 1/ \sqrt{n}
$
, where $n$ is the number of detectors. 
Therefore, this method provides a performance gain equivalent to increasing the Tibet-III array, which has approximately 600 detectors, 
by approximately 150 to 200 detectors. 
Improving the angular resolution through analysis alone, without increasing the number of detectors, 
is highly beneficial from both budgetary and operational perspectives.
We found that the combined-analysis method provides better direction determination even when the shower core is offset from the array center.
We propose the following explanations for this finding.
Although the cone-fit method determines the direction by fitting the shower surface using the position of the air-shower core obtained from the reconstruction analysis in Section~\ref{subsec:reconstruction},
the combined-analysis method is believed to incorporate and correct for secondary particle density distributions and hit-timing information.
This explanation suggests that introducing an effect similar to that of the combined-analysis method, which corrects for such effects, into the cone-fit method would lead to improved direction determination.
\begin{acknowledgements}
 This work is supported by the joint research program of the Institute for Cosmic Ray Research (ICRR), the
University of Tokyo, and the computation was carried out on the computer resource of the ICRR. M Takita,
M Ohnishi, and Y Katayose were also supported by JSPS KAKENHI Grant Number 22H04912. M Ohnishi
was supported by JSPS KAKENHI Grant Number 21H04464. 
We thank the members of Tibet AS$\gamma$ and
ALPACA for their helpful discussions.
\end{acknowledgements}

%
\printbibliography 
%
%
 %
%
 %
%
 %
 %
 %

 %
 
%
 %
 %
%
%
 %
 %
%
%
%

\end{document}